\documentclass[prr, 10pt, twocolumn, tightenlines, longbibliography, aps,superscriptaddress,nofootinbib]{revtex4-2}

\usepackage[utf8]{inputenc}
\usepackage[T1]{fontenc}
\usepackage{float}
\usepackage[english]{babel}
\usepackage[normalem]{ulem}
\usepackage{mathtools}
\usepackage{bbold}
\usepackage{graphicx}
\usepackage{multirow}
\usepackage{tabularx}
\usepackage[dvipsnames, table]{xcolor}
\usepackage{soul}
\usepackage[colorlinks=true,allcolors=blue]{hyperref}
\usepackage{amsmath}

\DeclarePairedDelimiter\ket{\lvert}{\rangle}

\DeclarePairedDelimiterX\braket[2]{\langle}{\rangle}{#1 \delimsize\vert #2}

\begin{document}
	
\title{Effective two- and three-body interactions between dressed impurities\\ in a tilted double-well potential}

		\author{F. Theel}
		\affiliation{Center for Optical Quantum Technologies, University of Hamburg, Department of Physics, Luruper Chaussee 149, D-22761, Hamburg, Germany}
        \author{A. G. Volosniev}
		\affiliation{Center for Complex Quantum Systems, Department of Physics and Astronomy,
Aarhus University, Ny Munkegade 120, DK-8000 Aarhus C, Denmark}
		\author{D. Diplaris}
		\affiliation{Center for Optical Quantum Technologies, University of Hamburg, Department of Physics, Luruper Chaussee 149, D-22761, Hamburg, Germany}
        \affiliation{The Hamburg Centre for Ultrafast Imaging, University of Hamburg, Luruper Chaussee 149, D-22761, Hamburg, Germany}
		\author{F. Brauneis}
		\affiliation{Department of Physics, Technische Universit\"{a}t Darmstadt, 64289 Darmstadt, Germany}
        \author{S. I. Mistakidis}
        \affiliation{Department of Physics, Missouri University of Science and Technology, Rolla, MO 65409, USA}
		\author{P. Schmelcher}
		\affiliation{Center for Optical Quantum Technologies, University of Hamburg, Department of Physics, Luruper Chaussee 149, D-22761, Hamburg, Germany}
		\affiliation{The Hamburg Centre for Ultrafast Imaging, University of Hamburg, Luruper Chaussee 149, D-22761, Hamburg, Germany}
	
\date{\today}

\begin{abstract} 
We explore the impact and scaling of effective interactions between two and three impurity atoms, induced by a bosonic medium, on their density distributions.  
To facilitate the detection of mediated interactions, we propose a setup where impurities are trapped in a tilted double-well potential, while the medium is confined to a ring. 
The tilt of the potential breaks the spatial inversion symmetry allowing us to exploit the population of the energetically elevated well as a probe of induced interactions. For two impurities, the interaction with the medium reduces the impurity population at the energetically elevated well, which we interpret as evidence of induced impurity-impurity attraction. Furthermore, the impact of an induced three-body interaction is unveiled by comparing the predictions of an effective three-body model with many-body simulations. We extend our study for induced interactions to a three-component mixture containing distinguishable impurities. Our results suggest pathways to detect and tune induced two- and three-body interactions.
\end{abstract}

\maketitle

\section{Introduction}
An impurity in a quantum medium is a crucial model for understanding polarized many-body systems, with the polaron quasiparticle offering a theoretical framework for its description~\cite{landau1933uber,pekar1946autolocalization}. The state-of-the-art approach for testing this framework and going beyond it, is based on quantum simulators that can be realized in cold-atom laboratories~\cite{Chevy2010,massignan2014,scazza2022,grusdt2025review,massignan2025polaron}. They permit elaborated investigations of both static~\cite{schirotzek2009,hu2016,jorgensen2016} and dynamical~\cite{cetina2015,Cetina2016,skou2021} properties of an impurity 
in three-dimensional systems. Furthermore, they allow to assess more exotic one-dimensional geometries~\cite{bloch2008} and the associated impurity physics~\cite{catani2012,Wenz2013}.

Systems involving more than a single impurity provide insights into the phenomenon of medium-induced correlations between dressed particles~\cite{charalambous2019,Paredes2024}. This is anticipated to be especially important in one spatial dimension (1D) where the role of interactions is often considered enhanced in comparison to higher dimensions~\cite{Kuramoto2020}. Theoretical modeling clearly shows that induced attraction~\cite{recati2005,Bruderer2007, schecter2014,brauneis2021,petkovic2022} lowers the energy and leads to clustering of two impurities in 1D in free space~\cite{Huber2019,will2021,Guo2024}, in a lattice potential~\cite{Dutta2013,pasek2019,keiler2020,Isaule2024}, and in a harmonic trap~\cite{dehkharghani2018,mistakidis2019,mistakidis2020b,mistakidis2020a,AnhTai2024}.

One strategy for observing the effect of weak mediated interactions is to use easily accessible one-body observables. To this end, the system is brought close to a transition point, where even slight perturbations can lead to dramatic effects~\cite{DeSalvo2019}. Ideally, the measurement should rely on the density of the impurity cloud, which is a routinely available observable experimentally. 
The transition point implies a certain energy landscape that can be simulated in cold-atom experiments by tailoring, for instance, an external potential~\cite{Gaunt_uniform,Folman_chips,kruger2010weakly}.

In this work, we propose an arguably simple design of a 1D Bose gas with a few impurities in such an energy landscape: The impurities being trapped in a tilted double-well potential are coupled to a Bose gas, which is confined to a ring potential. The tilt of the potential is an experimentally available knob~\cite{folling2007direct,levy2007ac} that can introduce a small energy scale into the problem -- the energy gap between the two minima of the potential. To investigate this model, we focus on a few-body system. These systems are of particular  interest~\cite{sowinski2019,Minguzzi2022,mistakidis2023} because they allow for accurate numerical solutions and for studying the emergence of many-body concepts, such as the medium-induced interactions, from the underlying microscopic physics.

The numerical investigation of the ground-state properties is performed using the \textit{ab-initio} Multi-Layer Multi-Configuration Time Dependent Hartree method for atomic mixtures (ML-MCTDHX)~\cite{cao2017,cao2013, kronke2013}. 
This grants access to mixtures consisting of two to three distinguishable and indistinguishable impurities immersed in a Bose gas. 
It is found that the density population of impurities at the energetically higher double-well site is sensitive to variations of the impurity-medium interaction strength. 
To interpret this observation, we devise two- and three-body effective models characterized by suitable two- and three-body induced contact interaction contributions. Although, the parameters of these models are determined by fitting to the energies of the many-body system, they also capture other observables -- such as densities -- either qualitatively or, for specific system sizes, even quantitatively. 
Our analysis confirms the importance of two-body effective interactions. 
The lesser-known three-body induced interactions play a less significant role in the regime of weak interactions, where the concept of induced interactions is most useful \cite{johnson2009, hammer2013, panochko2021, tajima2025}. This aligns with our expectations. In all cases, the scaling of the induced two- and three-body interactions is numerically extracted and corroborated by perturbation theoretic arguments.

This work unfolds as follows. In Section~\ref{sec:setup} we introduce our multicomponent setup and the concept of induced interactions. 
Section~\ref{sec:ML-X} discusses the key ingredients of the employed many-body variational method. 
Impurity-impurity induced interactions along with the effective two- and three-body models are analyzed in Sec.~\ref{sec:two_impurities_BB} for two indistinguishable bosonic impurities and in Sec.~\ref{sec:three_impurities_BBB} for three impurities.
Generalizations of our results to a three-component system are provided in Sec.~\ref{three_component_induced}. 
We conclude and discuss future extensions of our findings in Sec.~\ref{conclusions}. 

Additional technical details are presented in five appendices. 
Appendix~\ref{app:effective_model} explicates the exact diagonalization method used for the effective models. 
Appendix~\ref{app:single_impurity_effmass} discusses the localization behavior of a single impurity. Appendix~\ref{app:two_impurities_effmass} reveals the role of the quasi-particle effective mass in single-particle observables, while in Appendix~\ref{app:renorm_3body_int} we provide further numerical evidences corroborating the presence of effective three-body interactions.
Appendix \ref{app:impact_of_corr} focuses on the impact of correlations on different observables. In Appendix~\ref{app:scaling_behavior}, we examine the impact of finite-size effects that are intrinsic to our numerical analysis.
Finally, Appendix~\ref{app:diverging3-body} elaborates on the inherent logarithmic divergent behavior of the three-body interaction term and how it is circumvented.

\begin{figure*}
\centering
\includegraphics[width=0.9\linewidth]{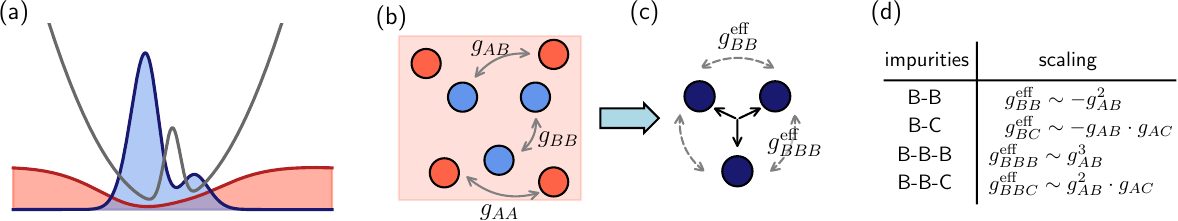}
\caption{Sketch of our impurity-medium setup. (a) One-body densities of the medium $A$ (red) and the three bosonic impurities (blue) for repulsive impurity-medium interactions. The medium, which consists of $N_A=12$ weakly interacting bosonic particles, is confined to a ring with periodic boundary conditions. The impurities are trapped by a tilted double-well potential (gray line). (b) The medium particles (red circles) interact via a contact interaction of strength $g_{AA}$. The impurities (blue circles) repel each other with strength $g_{BB}$. The boson-impurity interaction is denoted by $g_{AB}$. (c) The induced interactions between the three impurities are parameterized by the effective two-body ($g_{BB}^{\mathrm{eff}}$) as well as three-body ($g_{BBB}^{\mathrm{eff}}$) effective couplings. (d) Overview of the identified scaling behavior of the mediated two- and three-body interaction strengths between two or three impurities belonging to species $B$ and $C$.} 
\label{fig:setup_sketch}
\end{figure*}

\section{Multicomponent setup and induced interactions}
\label{sec:setup}

\subsection{Impurity-in-a-medium setting}

To study the induced two- and three-body interactions, we consider a bosonic medium consisting of $N_A$ atoms on a ring. The mass of a boson is $m_A$; the boson-boson interaction is parametrized by the standard contact interaction potential of strength $g_{AA}$~\cite{olshanii1998}. In the other component, we have up to three bosonic impurities with mass $m_B$ confined in a tilted 1D double-well potential \cite{dounas-frazer2007, carr2010, erdmann2018}. A free-space impurity-impurity contact interaction potential of strength $g_{BB}$ is assumed. The coupling strength is experimentally adjustable via either Fano-Feshbach tuning that changes the three-dimensional $s$-wave scattering length~\cite{chin2010} or by modifying the transverse confinement~\cite{Dunjko2011}. The later is assumed to be so tight that the transverse excitations are frozen out as in typical quasi-1D experiments, see, e.g., Refs.~\cite{bersano2018, lannig2020, romero-ros2024}.

The corresponding many-body Hamiltonian has the form,
\begin{align}
	\hat{H} = \hat{H}_A + \hat{H}_B + \hat{H}_{AB},
	\label{eq:MB_hamiltionian_AB}
\end{align}
where $\hat{H}_\sigma$ denotes the Hamiltonian of component $\sigma=\{A,B\}$ and $\hat{H}_{AB}$ represents the intercomponent interaction of effective strength $g_{AB}$. 
Specifically,
\begin{subequations}
\begin{align}
	\hat{H}_A &= -\sum_{i=1}^{N_A}
    \frac{\hbar^2}{2m_A} \partial_x^2 + g_{AA} \sum_{i<j} \delta(x_i^A-x_j^A), \\
	\hat{H}_B &= \sum_{i=1}^{N_B} \hat{h}_B^{(1)}(x_i^B) + g_{BB} \sum_{i<j} \delta(x_i^B-x_j^B),\label{bath_Hamilt}
	\\ 
	\hat{H}_{AB} &= g_{AB} \sum_{i=1}^{N_A} \sum_{j=1}^{N_B}\delta(x_i^A-x_j^B).
\end{align}
\end{subequations}
The Hamiltonian $\hat{h}_B^{(1)}(x_i^B) = - \frac{\hbar^2 }{2m_B }\partial_x^2 + V_B(x_i^B)$ describes a single impurity in a tilted double-well external trap. The latter is modeled by a superposition of a harmonic oscillator with frequency $\omega$, a Gaussian potential of width $w$ and height $h$, and a linear tilting potential of strength $\alpha$,  
\begin{align}
V_{B}(x) = \frac{1}{2} m_B \omega x^2 + \frac{h}{w \sqrt{2\pi}} \exp\left( -\frac{x^2}{2w^2} \right) + \alpha x.
\label{eq:double_well_potential}
\end{align}
It is illustrated in Fig.~\ref{fig:setup_sketch}(a) along with characteristic density distributions of the medium $A$ and the three impurities for repulsive impurity-medium couplings.
The presence of a small $\alpha$ breaks the inversion symmetry of the problem and leads to an energy offset between the two double-well sites. 
The tilted double-well can be readily implemented in experiments by imposing a bias potential~\cite{folling2007direct}, while the ring trap is realized using time-averaged potentials \cite{sherlock2011, bell2016}.

For simplicity, we study a mass-balanced mixture, namely it holds that $m_A=m_B\equiv m=1$, and employ harmonic oscillator units. Accordingly, the energy scales are expressed in units of $\hbar \omega$, while the length and interaction scales are in terms of $\sqrt{\hbar/m\omega}$ and $\sqrt{\hbar^3 \omega/m}$, respectively. Typically, our bath component consists of $N_A=12$ bosonic particles featuring ``weak'' intracomponent repulsion, $g_{AA}=g_{BB}=0.1$. By varying the parameter $g_{AB}$, we explore the strength of induced impurity-impurity interactions.
In our numerical simulations all atoms of the medium reside on a ring of length $L=12\sqrt{\hbar/m\omega}$ (the pre-factor here is given by $N_A$) ensuring that $L\gg\sqrt{\hbar/(m\omega)}$. This requirement reduces the role of finite-size effects in our study. 
Finally, throughout this work we employ a double-well characterized by $\omega=1$, $w=0.3 \sqrt{\hbar/m\omega}$ and $h=3\sqrt{\hbar m \omega}$, while the energy offset parameter is considered to be $\alpha=0.06 \sqrt{\hbar m \omega}$.
These are representative parameters which we employ for numerically solving the system under consideration and allow us to demonstrate the effect of induced interactions. The presence of the latter is independent of the size of the system, see Appendix \ref{app:scaling_behavior} for a brief discussion of the impact of the system size on our results.

Our multicomponent systems can be experimentally implemented, for instance, with different hyperfine states of a $^{87}$Rb gas. As an example the impurities may be realized using the state $|F = 1, m_F = 1 \rangle$ while the bosons are in the $|F = 2, m_F = 1 \rangle$ state~\cite{egorov2013}.
For the three-component mass-balanced system discussed in Section~\ref{sec:two_impurities_BC}, it is possible to utilize an additional hyperfine state, e.g., $|F=1, m_F=-1 \rangle$ of $^{87}$Rb~\cite{bersano2018}. 

Within this work we devise two- and three-body models which employ effective two- and three-body interaction parameters to effectively describe the behavior of the impurities interacting with the majority species, see Figs. \ref{fig:setup_sketch}(b) and (c) for a sketch.

\subsection{Induced interactions}
\label{sec:pert_induced}

As the focus of this paper is induced interactions, we briefly introduce this concept here, first for a system with two impurities. Our results are applicable to any external trapping of the impurities as long as the bosonic medium is confined to a ring potential of length $L$.
We utilize perturbation theory to calculate the correction to the non-interacting energy due to interactions~\cite{Landau1981-ra}
\begin{equation}
\delta E = M_{gg} + \sum_{e} \frac{ | M_{eg} |^2} {E_g - E_e} +...\;,
\label{eq:perturbation_theory}
\end{equation}
where $M_{ij}=\langle i| \hat H_{AB}|j\rangle $ is the non-interacting matrix element between the states $i$ and $j$; the index $g$ ($e$) denotes the ground (excited) state of the non-interacting system characterized by the energy $E_g$ ($E_e$). Using indistinguishability of particles, we write the matrix element in the coordinate space representation as 
\begin{equation}
M_{eg}= g_{AB} N_A N_B \int\mathrm{d}x_{1}^A\mathrm{d}x_{1}^B \Psi_{g} \delta(x_1^A-x_1^B) \Psi_e .
\label{eq:matrix_element}
\end{equation}

Induced interactions is an intuitive method to interpret the energy difference $\delta E_2= \delta E(N_B=2)-2\delta E(N_B=1)$, which is in general non-zero. To demonstrate this, note that for $N_B=1$ [$N_B=2$] the non-interacting ground state can be written as:
$\Psi_g=\phi_g(x^B_1)\prod_i \psi_g(x_i^A)$ [$\Psi_g=\phi_g(x^B_1)\phi_g(x^B_2)\prod_i \psi_g(x_i^A)$],
where we assume that all bosons occupy the same orbital, $\psi_g$; $\phi_g$ is the ground state of $\hat{h}_B^{(1)}$. For bosons on a ring, it holds that $\psi_g(x_i^B)=1/\sqrt{L}$. 
Furthermore, we consider only excitations of the medium, as these are essential for induced interactions. The corresponding excited states read: 
$\Psi_e=\phi_g(x_1^B)\sum_i \psi_{e}(x_i^A)/\sqrt{N_A L^{N_A-1}}$ [$\Psi_e=\phi_g(x_1^B)\phi_g(x_2^B)\sum_i \psi_{e}(x_i^A)/\sqrt{N_A L^{N_A-1}}$]. 
Using these expressions in Eq.~(\ref{eq:matrix_element}) the energy difference becomes 
\begin{equation}
\delta E_2\simeq  2 g_{AB}^2 \frac{N_A}{L} \int \mathrm{d}x_1^{B}\mathrm{d}x_2^{B} |\phi_g(x_1^{B})|^2 |\phi_g(x_2^{B})|^2 V_{II},
\label{eq:deltaE2_for_two_particles}
\end{equation} 
with the function $V_{II}$ defined as follows 
\begin{equation}
V_{II}(x_1^{B},x_2^{B})= \sum_{e}\frac{\psi^*_e(x_2^{B})\psi_e(x_1^{B})}{E_g - E_e}.\label{eq:perturb_inter}
\end{equation}
Note that the expression in Eq.~(\ref{eq:deltaE2_for_two_particles}) is equivalent to the first-order perturbative correction to the energy of two non-interacting impurities assuming that $V_{II}$ is a perturbation. The fact that $V_{II}$ does not depend on the state of the impurity enforces the interpretation of $V_{II}$ in terms of an effective two-body interaction.  Note that the leading-order contribution to the energy from the two-body induced interaction is proportional to $g_{AB}^2$, see also Fig.~\ref{fig:setup_sketch}(d). It is always attractive as it stems from second-order perturbation theory.

We remark that Eq.~(\ref{eq:perturb_inter}) is also the correction to the non-interacting Bose gas perturbed by two static impurity potentials located at $x_1^{B}$ and $x_2^{B}$. This allows one to calculate $V_{II}$ using the theoretical methods presented in Refs.~\cite{recati2005,mistakidis2020b,brauneis2021,petkovic2022,will2021} (see also  Refs.~\cite{naidon2018,Panochko2022,Jager2022,Drescher2023} for relevant works in higher dimensions). In particular, one can approximate\footnote{Note that here we disregard the long-range part of the potential~\cite{schecter2014,reichert2019a}, which is irrelevant for small trapped systems~\cite{mistakidis2020b}.} $V_{II}\simeq -\delta(x_1^{B}-x_2^{B}) L/(2 g_{AA} N_A)$ for weak interactions in the thermodynamic limit ($N_A\to\infty, L\to\infty$ and $N_A/L = \mathrm{const}$)~\cite{mistakidis2023}.

For three impurities, one introduces a three-body effective interaction to interpret the energy difference $\delta E_3=\delta E(N_B=3)-3\delta E_2(N_B=2)-3\delta E(N_B=1)$. Here, the factor $3$ in front of $\delta E_2(N_B=2)$ accounts for the number of interacting pairs. Alternatively, one can think that this factor is chosen so that $\delta E_3$ vanishes at the level of second order perturbation theory. Therefore, it is necessary to consider the energy correction within third-order perturbation theory
\begin{equation}
\delta_3= \sum_{e \neq e'} \frac{M_{ge} M_{ee'}M_{e'g}}{\left(E_g - E_e \right) \left(E_g - E_e' \right)} -  \sum_{e} \frac{M_{gg}|M_{eg}|^2}{\left( E_g - E_e \right)^2}.
\label{eq:delta3}
\end{equation}
We shall only analyze overall features of this expression. To this end, we consider the following excited states
$\Psi_e=\phi_g(x_1^B)\phi_g(x_2^B)\phi_g(x_3^B)\sum_i\psi_{e}(x_i^A)/\sqrt{N_A L^{N_A-1}}$, which are motivated by our discussion of $V_{II}$.  
Using these states, we write the contribution to the energy due to the first term in Eq.~(\ref{eq:delta3}) that can be interpreted as a result of three-body induced interactions: $\frac{6 g_{AB}^3 N_A}{L} \int \mathrm{d}x_1^{B}\mathrm{d}x_2^{B} \mathrm{d}x_3^{B} |\phi_g(x_1^{B})|^2 |\phi_g(x_2^{B})|^2 |\phi_g(x_3^{B})|^2 V_{III}$,
where 
\begin{equation}
V_{III}(x_1^{B},x_2^{B},x_3^{B})= \sum_{e,e'} \frac{\psi_e(x_1^{B})\psi_e^*(x_2^{B})\psi_{e'}^*(x_3^{B})\psi_{e'}(x_2^{B})}{\left(E_g - E_e \right) \left(E_g - E_e' \right)}.
\label{eq:VIII}
\end{equation} 
Note that $V_{III}$ is a product of two-body interactions, i.e., $V_{III}(x_1^{B},x_2^{B},x_3^{B})=V_{II}(x_1^{B},x_2^{B}) V_{II}(x_2^{B},x_3^{B})$. For weak interactions in the thermodynamic limit, we can therefore write $V_{III}(x_1^{B},x_2^{B},x_3^{B})=\delta(x_1^{B}-x_2^{B})\delta(x_2^{B}-x_3^{B})L^2/(4g_{AA}^2 N_A^2)$. 
As the three-body induced interactions is proportional to $g_{AB}^3$ in the leading order, we expect it to be repulsive for $g_{AB}>0$ and attractive for $g_{AB}<0$, see also Fig~\ref{fig:setup_sketch}(d).
Note that the last term in Eq.~(\ref{eq:delta3}) gives rise to an effective two-body interaction in $g_{AB}^3$-order that depends on the number of impurities.   
Having introduced the general features of induced interactions, we explicate them further in Sections \ref{sec:two_impurities_BB} and \ref{sec:three_impurities_BBB} for the cases with $N_B=2$ and $N_B=3$, respectively.

\section{Many-body approach}
\label{sec:ML-X}

To study the ground state properties of our quantum many-body system, we employ the \textit{ab-initio} ML-MCTDHX method~\cite{cao2017,cao2013,kronke2013}. 
A main facet of this approach is that the full many-body wave function is expressed in a multi-layer structure with time-dependent and variationally optimized basis functions. 
This process is tailored to account for the relevant intra- and intercomponent correlations of multicomponent cold atom settings. 
Detailed discussions on the ingredients, successful applicability and reductions of this method for a plethora of cold atom systems can be found in the recent reviews~\cite{Lode_review,mistakidis2023}.

Below, we mainly elaborate on the structure of the many-body wave function for the most general three-component setting used in our analysis. Comments on the reduction of this scheme to the two-component setup are provided whenever appropriate, see also~\cite{keiler2021,theel2024} for more detailed discussions. 
For a three-component system the wave function is firstly expanded in the truncated basis comprising of $D_{\sigma}$, with $\sigma= A,B,C$, orthonormal time-dependent species functions, $\ket{\Psi^{\sigma}_i(t)}$, as follows
\begin{align}
    | \Psi^{MB} (t) \rangle = \sum_i^{D_A} \sum_j^{D_B} \sum_k^{D_C} A_{ijk}(t) |\Psi_i^A (t) \rangle |\Psi_j^B (t) \rangle |\Psi_k^C (t) \rangle.
    \label{eq:psi_mlx}
\end{align}
Here, $A_{ijk}(t)$ represent the time-dependent expansion coefficients. This expansion grants access to intercomponent correlations. Specifically, the expansion coefficients referring to the contribution of each species function provide information about the intercomponent entanglement, see also Refs.~\cite{horodecki2009}, since they allow the evaluation of the eigenvalues of the species reduced density matrices~\cite{vidal2002, plenio2005, theel2024}.
In the case of a binary mixture, the above expansion reduces to a truncated Schmidt decomposition~\cite{schmidt1907, ekert1995}, see for instance the works~\cite{mistakidis2019c, theel2020, pyzh2022} and references therein.

Next, in order to incorporate intracomponent correlations into our ansatz, each of the species functions, $|\Psi_i^{\sigma} (t) \rangle$, is expanded in terms of the bosonic number states $|\vec{n}^{\sigma}_t\rangle$. The latter are weighted by the time-dependent coefficients $C_{i,\vec{n}^{\sigma}}^{\sigma} (t)$. This yields 
\begin{align}
    |\Psi^{\sigma}_i (t) \rangle = \sum_{\vec{n}|N_{\sigma}}  C_{i,\vec{n}^{\sigma}}^{\sigma} (t) |\vec{n}^{\sigma} (t)\rangle, 
\end{align}
where $N_{\sigma}$ bosons are allowed to occupy $d_{\sigma}$ single-particle functions (SPFs) $|\phi_j^{\sigma} (t) \rangle$.  The vector $\vec{n}^{\sigma} = (n_1^{\sigma}, \dots, n_{d_{\sigma}}^{\sigma})$ 
indicates the occupation number of each SPF. Finally, the SPFs are expanded with respect to a time-independent basis consisting of $\mathcal{M}_{pr}$ grid points\footnote{For a given $\mathcal{M}_{pr}$, ML-MCTDHX is numerically exact when $d_{\sigma}=\mathcal{M}_{pr}$ and  $D_{\sigma}$ equals the number of bosonic configurations, i.e. $\binom{N_{\sigma} + d_{\sigma} - 1}{d_{\sigma} - 1}$.}.
The ML-MCTDHX equations of motion for the above-described coefficients are derived, e.g., by using the Dirac-Frenkel variational principle $\langle \delta \Psi | (i \hbar \partial_t - \hat{H} ) | \Psi \rangle = 0$.
A limiting case is to set $D_A = D_B = D_C = 1$, which leads to a single product state in Eq.~(\ref{eq:psi_mlx}) neglecting  intercomponent correlations, but still including intracomponent ones.
In addition, using $d_A = d_B =d_C = 1$, the method reduces to the standard mean-field  approach where all correlations are absent. 
We will exploit in Appendix~\ref{app:impact_of_corr} different %types of sMF 
reduction ansatzes in order to unravel the impact of two- and three-component correlations on one- and two-body observables.

%========================================================================================
% B-B system
%========================================================================================
\section{Two bosonic impurities}
\label{sec:two_impurities_BB}

We start our investigation on induced interactions with a system containing two bosonic impurities, i.e., $N_B=2$. In what follows, the many-body ML-MCTDHX computations of the corresponding impurity-medium setting are analyzed and subsequently compared with an effective two-body model and the standard mean-field approximation.

\begin{figure}
\centering
\includegraphics[width=\linewidth]{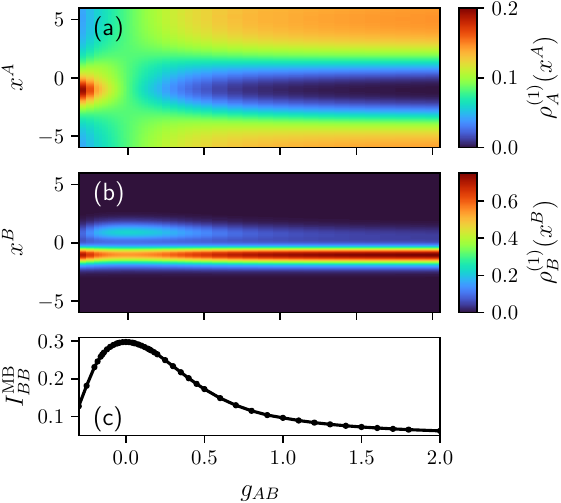}
\caption{Ground-state one-body densities of (a) the medium and (b) the two bosonic impurities as a function of the interspecies interaction strength $g_{AB}$. 
(c) Population of impurities at the energetically elevated double-well site  (located at $x^B>0$), $I_{BB}^{\mathrm{MB}}$, see Eq. (\ref{eq:int_obd}) for the definition, with respect to $g_{AB}$. The simulations are performed within the many-body approach ML-MCTDHX. 
The two repulsively interacting ($g_{BB}=0.1$) impurities experience a tilted double-well potential and are coupled to a bosonic medium with $N_A=12$ and $g_{AA}=0.1$.}
\label{fig:BB_results_1bd}
\end{figure}

\subsection{One-body density configurations}

By choosing to work with a tilted double-well potential, we intentionally break the system's inversion symmetry. This creates an energy offset between the two wells of the double-well potential. Consequently, the impurities prefer to occupy the energetically lower well.
In Figs. \ref{fig:BB_results_1bd}(a) and (b) we present the one-body densities $\rho_\sigma^{(1)}(x^\sigma)$ of species $\sigma=A,B$, respectively, as a function of the impurity-medium interaction strength $g_{AB}$, for fixed $g_{AA}=g_{BB}=0.1$.
Our choice of $g_{BB}$ is somewhat arbitrary; we select it to match the value of $g_{AA}$. Note that for the induced interactions to exist in the thermodynamic limit, $g_{AA}$ should be finite, see Sec.~\ref{sec:pert_induced}. 

As it can be readily seen from Fig. \ref{fig:BB_results_1bd}(b), the population imbalance of the impurities, i.e., the imbalance of the impurities' one-body density with respect to the two double-well sites, is evident already in the non-interacting case, $g_{AB}=0$. This behavior becomes gradually more prominent with increasing intercomponent interaction $|g_{AB}|$ leading in the strongly interacting case ($g_{AA}, g_{BB} \ll g_{AB}$) to a depopulation of the energetically higher site from the impurities.
The density of the medium is reduced (increased) for $g_{AB}>0$ ($g_{AB}<0$), as shown in Fig.~\ref{fig:BB_results_1bd}(a), at the location of the impurities.
Indeed, as long as $g_{AB}>0$ the medium atoms prefer to avoid the impurities, while if $g_{AB}<0$ the bosons accumulate in the vicinity of the impurities.

To quantify the depopulation process of the impurities we integrate $\rho_B^{(1)}(x^B)$ over the energetically higher double-well site (located at $x^B>0$):
\begin{align}
I_{BB} = \int_0^{L/2} \rho_B^{(1)}(x^B) \mathrm{d}x^B,
\label{eq:int_obd}
\end{align} 
where the upper integration limit is set by the length $L$ of the ring potential. 
A typical profile of this quantity is presented in Fig.~\ref{fig:BB_results_1bd}(c) with respect to $g_{AB}$ and labeled as $I_{BB}^{\mathrm{MB}}$ to indicate that this observable has been obtained within the full many-body approach (cf. Section \ref{sec:ML-X}). Here, the largest population at the energetically higher site occurs at $g_{AB}=0$ and then reduces for finite values of $g_{AB}$, thus implying that the impurities move to the energetically lower site at $x^B<0$. For a detailed discussion of the impact of the system size on $I_{BB}$ see Appendix \ref{app:scaling_behavior}.

In the following our goal is to construct an effective model that captures the above-discussed depopulation mechanism. Thereby, we construct an effective two-body model whose parameters are determined on the basis of the polaron and bipolaron energies. [In this paper, we use the terms `polaron' and `bipolaron' to refer to a Bose gas with a single and two impurities, respectively. This terminology became standard, even when working with a few-body system~\cite{mistakidis2019,AnhTai2024,Isaule2024}.]
To validate this effective model, we compare its predictions regarding the integrated density with those of the many-body approach, $I_{BB}^{\mathrm{MB}}$, and of the mean-field approximation, $I_{BB}^{\mathrm{MF}}$.

\begin{figure*}
\centering
\includegraphics[width=\linewidth]{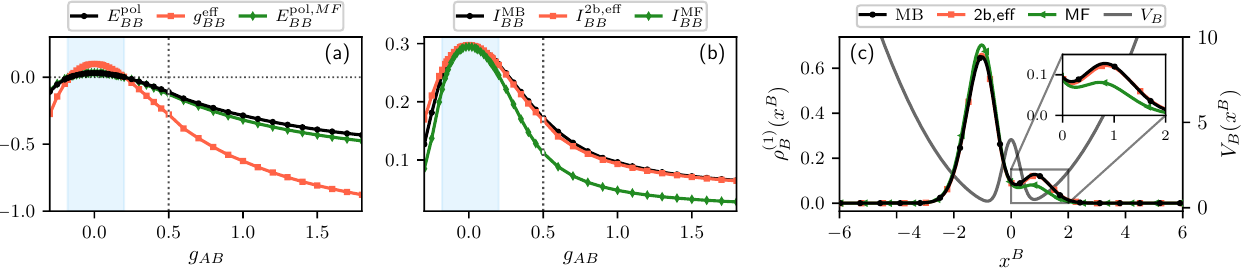}
\caption{(a) The energy $E_{BB}^{\mathrm{pol}}$ (in units of $\hbar \omega$) and the corresponding effective interaction strength $g_{BB}^{\mathrm{eff}}$ (in units of $\sqrt{\hbar^3 \omega/m}$) in terms of $g_{AB}$.
The energies are obtained either using the many-body approach, $E_{BB}^{\mathrm{pol}}$, or the mean-field approximation, $E_{BB}^{\mathrm{pol, MF}}$ (see legend). 
Notice that the predictions of the two approaches agree well. 
(b) Integrated one-body density [Eq. (\ref{eq:int_obd})] determined via the many-body approach $I_{BB}^{\mathrm{MB}}$, the effective two-body model [Eq. (\ref{eq:eff_ham_BB})], $I_{BB}^{\mathrm{eff}}$, and the mean-field approximation, $I_{BB}^{\mathrm{MF}}$. The blue shaded areas in panels (a) and (b) mark the interaction region where the relative deviation between $I_{BB}^{\mathrm{MB}}$ and $I_{BB}^{\mathrm{MF}}$ is smaller than $0.1$. (c) One-body density of the impurities within the many-body approach (MB), the effective two-body model (2b,eff) and the mean-field approximation (MF), see also legend. The double-well potential is also shown and the inset illustrates the impurities' density within the right well as predicted for the different methods. Here, $g_{AB}=0.5$, see the vertical dotted lines in panels (a) and (b). 
Other system parameters are $N_A=12$ and $g_{AA}=g_{BB}=0.1$.}
\label{fig:BB_results_fit}
\end{figure*}

\subsection{Effective two-body model}
\label{sec:BB_eff_2b_model}

In the previous section we have seen that finite impurity-medium coupling strengths $g_{AB}$ lead to a reduction of the impurities' density at the energetically higher double-well site.
There are two plausible mechanisms for this behavior. First, it may be attributed to the increase of the effective mass of the impurity, which leads to spatial localization, see also Appendix~\ref{app:single_impurity_effmass}. Second, this behavior can be interpreted as an additional mediated interaction between the impurities induced by their coupling ($g_{AB}$) with the medium.

To construct an effective two-body model, we first define the energy cost of adding the impurity to the system, $E_B^{\mathrm{pol}}$ \cite{grusdt2025review,massignan2025polaron},
\begin{align}
E_B &= E_A^{(0)} + E_B^{\mathrm{pol}}.
\label{eq:Epol_B}
\end{align} 
Here, $E_B$ is the total energy of the impurity-bath system, and $E_A^{(0)}$ denotes (throughout this work) the energy of the bath without any impurities. 
Note that $E_B$ and  therefore $E_B^{\mathrm{pol}}=E_B - E_A^{(0)}$  depend on $g_{AB}$. 
The total energy is used, together with the undisturbed impurity Hamiltonian $\hat{h}_B^{(1)}$ [Eq.~(\ref{bath_Hamilt})], to formulate the effective model 
\begin{align}
\hat{H}_{B}^{\rm{eff}}(x) =& \epsilon_B^{\mathrm{pol}} + \hat{h}_B^{(1)}(x),
\label{eq:eff_ham_B}
\end{align} 
where $\epsilon_B^{\mathrm{pol}} = E_B^{\mathrm{pol}} - E^{(1\rm{body})} $ is chosen such that the ground state energy of $\hat{H}_{B}^{\rm{eff}}$ matches the energy $E_B^{\mathrm{pol}}$.  $E^{(1\rm{body})}$ is the ground-state energy of the one-body Hamiltonian $\hat{h}^{(1)}_B(x)$. 
In this effective one-body model we explicitly consider the bare impurity mass $m_B$ as it turns out that for the considered parameters the effective two-body model predictions are improved in the absence of the effective mass for $g_{AB}>0$. More details about this fact can be found in Appendix \ref{app:single_impurity_effmass} where the effective mass is determined, and in Appendix \ref{app:two_impurities_effmass} at which the behavior of the induced two-body interaction accounting for the effective mass is reported.

Let us now consider two impurities. Each impurity contributes with $E^{\mathrm{pol}}_B$ to the energy of the undisturbed bath,  $E_A^{(0)}$. The total energy, $E_{BB}$,  of a system consisting of a bath coupled to two impurities can then be decomposed~\cite{camacho-guardian2018a, pasek2019, Isaule2024} (note the resemblance to the perturbative analysis presented in Sec.~\ref{sec:pert_induced}), as 
\begin{align}
E_{BB} &= E_A^{(0)} + 2E_B^{\mathrm{pol}} + E_{BB}^{\mathrm{pol}}. 
\label{eq:Epol_BB}
\end{align}
In this expression, $E_{BB}^{\mathrm{pol}}$ appears due to the direct and effective interactions between the impurity atoms. 
In Fig.~\ref{fig:BB_results_fit}(a), we present the many-body results of $E_{BB}^{\mathrm{pol}}$ for $g_{BB}=0.1$. For weak impurity-medium couplings, the positive value of $g_{BB}$ implies that $E_{BB}^{\mathrm{pol}}>0$. However, for $g_{AB}\gtrsim g_{AA}$, induced attractive interactions dominate~\cite{mistakidis2023}. In this regime $E_{BB}^{\mathrm{pol}}<0$, which suggests clustering of impurities\footnote{In the homogeneous case the induced interaction between two impurities is approximately given by $g_{BB}^{\mathrm{eff}} -g_{BB}\simeq - \frac{g_{AB}^2}{g_{AA}}$ \cite{mistakidis2023}, see also Section \ref{sec:pert_induced}. The induced attraction is roughly equal to the internal impurity-impurity repulsion ($g_{BB}=g_{BB}^{\mathrm{eff}}$) when $g_{AB}\simeq g_{AA}$. We have checked that the mediated interaction $g_{BB}^{\mathrm{eff}}$ determined via the fitting procedure indeed approaches this prediction if (i) $g_{AB} \ll g_{AA}$ and (ii) the healing length of the Bose gas is much smaller than $L$.}.
To capture this effect, we design an effective two-body model that incorporates an effective interaction of strength $g_{BB}^{\mathrm{eff}}$ induced by the medium
\begin{align}
\hat{H}_{BB}^{\rm{eff}} =& \sum_{i=1}^{2} \hat{H}_{B}^{\rm{eff}}(x_i) + g_{BB}^{\mathrm{eff}}\delta(x_1-x_2).
\label{eq:eff_ham_BB}
\end{align}
For simplicity we consider here a contact effective interaction, which is motivated by the discussion in Sec.~\ref{sec:pert_induced}. We anticipate that a `simple' delta-function potential form cannot capture the physics of strong interactions in full detail \cite{dehkharghani2018, mistakidis2020b}. Still, as we shall argue below it provides an adequate starting approximation even in this case.

To calculate $g_{BB}^{\mathrm{eff}}$, the condition that the ground state energy, $E_{BB}^{{\mathrm{eff}}}$, of $\hat{H}_{BB}^{\rm{eff}}$ matches our many-body results is imposed, namely
\begin{align}
E_{BB}^{{\mathrm{eff}}} \stackrel{!}{=} 2E_B^{\rm{pol}} + E_{BB}^{\mathrm{pol}}.
\label{eq:energy_condition_BB}
\end{align}
In practice, the free parameter $g_{BB}^{\mathrm{eff}}$ is varied until Eq.~(\ref{eq:energy_condition_BB}) is satisfied.
To solve the two-body model, and later on the three-body one, we expand the respective wave function in terms of number states which are composed of a set of static single-particle functions, see Appendix \ref{app:effective_model} for more details.

In Fig. \ref{fig:BB_results_fit}(a), $g_{BB}^{\mathrm{eff}}$ is illustrated for $g_{BB}=0.1$ as a function of $g_{AB}$. Apparently, for $g_{AB}=0$, induced impurity-impurity interactions are absent and it holds that $g_{BB}^{\mathrm{eff}}=g_{BB}$\footnote{In fact, the effective model at $g_{AB}=0$ yields a $g_{BB}^{\mathrm{eff}}$ which matches up to the third digit $g_{BB}=0.1$. This deviation marks the accuracy of the employed exact diagonalization method when fitting to the ML-MCTDHX data.}.
Although, the effective model was established using purely energy considerations, it also turns out to be useful for calculating other observables. This fact is shown in Fig. \ref{fig:BB_results_fit}(b), (c) where we compare the impurities integrated one-body densities and a characteristic spatial profile obtained with the effective two-body model, $I_{BB}^{\mathrm{2b,eff}}$, and the many-body approach, $I_{BB}^{\mathrm{MB}}$. An excellent agreement between the two methods is observed for the aforementioned population of the energetically higher double-well site and the density configuration itself. 
Non-negligible deviations occur only upon increasing the  impurity-medium attraction. The excellent agreement between the effective model and the many-body results at large positive values of $g_{AB}$ is coincidental. Indeed, there is no reason to expect our simple effective model to remain valid in the strong interaction regime. Recall that we approximate the two-body as well as the three-body induced interaction potential by a contact interaction potential. To further improve the agreement with the many-body results more elaborated potentials might be required, see also \cite{chen2018, dehkharghani2018}. To illustrate the breakdown of the effective model, we consider larger systems in Appendix \ref{app:scaling_behavior}.

To study the impact of correlations on the employed one-body observables, we perform a comparison of our results to the outcome of the standard mean-field approximation. 
Inspecting Figs. \ref{fig:BB_results_fit}(a) and~(b) it turns out that the mean-field results are accurate for weak intercomponent interactions. However, for larger values of $|g_{AB}|$ the mean-field predictions for the one-body density start to  deviate from the many-body results indicating an increasing role of correlations in the system, see in particular Figs.~\ref{fig:BB_results_fit}(b), (c). This deviation implies that it is possible to study the strength of induced impurity-impurity interactions by observing the one-body density. 
For convenience, the interaction region ($-0.18<g_{AB}<0.2$) where the mean-field treatment yields accurate results (within 10\% accuracy) for the integrated density is marked by a blue shaded area in Figs. \ref{fig:BB_results_fit}(a) and (b).
The origin of the deviation between the many-body and the mean-field results is linked to interspecies correlations, see Appendix~\ref{app:impact_of_corr} for more details.

Complementary, we have determined the natural orbitals $n_i^\sigma$ and natural populations $\phi_{\mathrm{nat},i}^{\sigma}(x)$ of species $\sigma$ by expressing the respective one-body density matrices in their spectral form $\rho_\sigma^{(1)}(x, x') = \sum_{i=1}^{d_{\sigma}} n_i^\sigma \phi_{\mathrm{nat},i}^{\sigma}(x) (\phi_{\mathrm{nat},i}^{\sigma}(x'))^*$ \cite{sakmann2008,mistakidis2018}.
We detect a small fragmentation of both  the majority ($A$) and the impurity ($B$) species, with the depletion of the largest populated natural orbital ranging from 
%being of the order of 
$1-n_1^{\sigma} \sim 10^{-3} $ to $ 10^{-2}$ for $|g_{AB}|\lesssim0.4$. Here, we note that,  naturally, the depletion of the impurities is somewhat larger than the one of the majority species.  These findings further indicate that the observed  deviations  between mean-field and many-body methods stem from interspecies entanglement.

%========================================================================================
% B-B-B system
%========================================================================================

\begin{figure*}
\centering
\includegraphics[width=\linewidth]{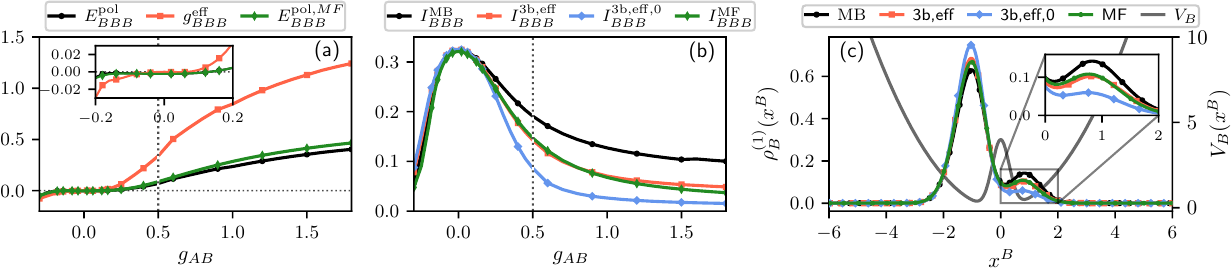}
\caption{(a) The energy $E_{BBB}^{\mathrm{pol}}$ and the effective three-impurity interaction strength $g_{BBB}^{\mathrm{eff}}$ calculated by fitting the effective model to $E_{BBB}^{\mathrm{pol}}$ (see main text). $E_{BBB}^{\mathrm{pol}}$ is compared to $E_{BBB}^{\mathrm{pol}, MF}$, which is computed within the mean-field approach. 
(b) Integrated one-body density of the impurities, $I_{BBB}$, computed within the many-body (MB) method, the mean-field (MF) approximation and the three-body effective model with and without a three-body interaction potential, labeled as (3b,eff) and (3b,eff,0), respectively. 
(c) Ground-state density distribution of the impurities within four different approaches (see legend) for fixed $g_{AB}=0.5$ indicated also by vertical gray dotted lines in panels (a) and (b). The inset provides a magnification of the densities for the right well of the double well (see also the right axis of panel (c)) emphasizing the degree of agreement among the different approaches. 
Other system parameters are $N_A=12$, $N_B=3$, $g_{AA}=0.1$ and $g_{BB}=0.1$.}
\label{fig:BBB_results_fit}
\end{figure*}

\section{Three bosonic impurities}
\label{sec:three_impurities_BBB}

Next, we investigate three bosonic impurities coupled to a bosonic medium. Besides an effective two-body interaction discussed in the previous section, the system can experience an effective three-body force (cf. Ref.~\cite{Guo2024}). To quantify its effect, it is necessary to extend the effective model given by Eq.~(\ref{eq:eff_ham_BB}). 

Similar to the two-impurity case, the extended effective model is developed using energy considerations. Namely, the total energy is re-arranged as follows 
\begin{align}
    E_{BBB} &= E_A^{(0)} + 3E_B^{\mathrm{pol}} + 3E_{BB}^{\mathrm{pol}} + E_{BBB}^{\mathrm{pol}}.
    \label{eq:Epol_BBB}
\end{align}
This expression includes the unperturbed bath energy $E_A^{(0)}$ and three times $E_B^{\mathrm{pol}}$. The higher-order contributions enter via the two-body polaron energy $E_{BB}^{\mathrm{pol}}$, and the three-polaron one, $E_{BBB}^{\mathrm{pol}}$.
The coefficient $3$ in front of $E_{BB}^{\mathrm{pol}}$ reflects the number of interacting impurity-impurity pairs. It also follows from the analysis based upon perturbation theory, see Section \ref{sec:pert_induced}.

We incorporate induced three-body interactions in an effective three-body Hamiltonian as follows\footnote{Note that our three-body interaction term leads to divergences similar to the two-dimensional contact interaction, see Ref.~\cite{Pricoupenko2019} and Appendix~\ref{app:diverging3-body}. We renormalize this interaction potential by fitting to a finite Hilbert space.}
\begin{align}
\hat{H}_{BBB}^{\mathrm{eff}} =& \sum_{i=1}^{3} H_{B}^{\mathrm{eff}}(x_i) +  g_{BB}^{\mathrm{eff}} \sum_{\substack{i,j=1 \\ i\neq j}}^{3} (x_i - x_j) \nonumber \\
+&  g_{BBB}^{\mathrm{eff}} \delta(x_1-x_2)\delta(x_2-x_3),
\label{eq:eff_ham_BBB}
\end{align}
where the shape of the last term is motivated by the discussion in Sec.~\ref{sec:pert_induced}.
The parameter $g_{BB}^{\mathrm{eff}}$ is determined within the effective two-body model introduced in Sec. \ref{sec:BB_eff_2b_model}.
To compute the strength of the effective three-body interaction, we enforce the condition that the ground-state energy of $\hat{H}_{BBB}^{\mathrm{eff}}$, namely $E_{BBB}^{\mathrm{eff}}$, matches the energy of the three dressed impurities,
\begin{align}
E_{BBB}^{{\mathrm{eff}}} \stackrel{!}{=} 3E_B^{\rm{pol}} + 3E_{BB}^{\mathrm{pol}} + E_{BBB}^{\mathrm{pol}}.
\label{eq:energy_condition_BBB}
\end{align}
As in Eq.~(\ref{eq:Epol_BBB}), the one-body term here refers to the single-polaron energy $E_B^{\rm{pol}}$. 
The two-body interaction term accounts for the impurity-impurity correlations and $E_{BBB}^{\mathrm{pol}}$ implies the presence of the three-impurity induced interactions.

The analysis of a three-impurity system is shown in Fig. \ref{fig:BBB_results_fit} for varying $g_{AB}$ and constant $g_{BB}=0.1$. For weak interactions (i.e. $g_{AB}\to 0$), $E_{BBB}^{\mathrm{pol}}$ and $g_{BBB}^{\mathrm{eff}}$ appear to scale as $g_{AB}^3$, see Fig.~\ref{fig:BBB_results_fit}~(a), in agreement with the perturbative predictions discussed in Sec.~\ref{sec:pert_induced}. Also, despite the fact that the effective model of Eq.~(\ref{eq:eff_ham_BBB}) is constructed through energy considerations, it turns out to be useful for other observables (see below).  

Figure \ref{fig:BBB_results_fit}(b) depicts the integrated one-body density using $I_{BBB}^{\mathrm{MB}}$, obtained from the many-body treatment, as a reference. First, we determine the integrated one-body density from the effective model in the absence of three-body effects, i.e., setting $g_{BBB}^{\mathrm{eff}}=0$ in Eq. (\ref{eq:eff_ham_BBB}) which we refer to as $I_{BBB}^{\mathrm{3b,eff,0}}$.  
It is found that $I_{BBB}^{\mathrm{3b, eff,0}}$ agrees well with $I_{BBB}^{\mathrm{MB}}$ only for small impurity-medium coupling strengths, and that $g_{BBB}^{\mathrm{eff}}\neq 0$ leads in general to more accurate results.
Indeed, for larger values of $g_{AB}$, the integrated density of the impurities in either of the effective models deviates from the many-body result as can be readily seen in Fig.~\ref{fig:BBB_results_fit}(b). 
This implies that the assumed effective interactions capture only approximately induced correlations between particles.
This conclusion is further supported by investigating the one-body density for $g_{AB}=0.5$, see Fig. \ref{fig:BBB_results_fit}(c) and its inset.

We conclude that while three-body effects are clearly present, their analysis appears to be more involved than the one of two impurities. 
To interpret this observation, note that according to Section~\ref{sec:pert_induced}, the interaction strength $g_{BB}^{\mathrm{eff}}$ should be modified to account for the presence of the third impurity.
However, this modification alone is not sufficient to explain our numerical data, see Appendix \ref{app:renorm_3body_int}.

Finally, it is worth noting that the mean-field approximation yields results for the one-body densities and the energy that are comparable to those of the effective model, but not to those of the full many-body approach, see Fig. \ref{fig:BBB_results_fit}(a)-(c). The lack of agreement between the mean-field and many-body approaches is not coincidental -- it persists even for larger system sizes indicating that inter-particle correlations have a significant impact on the one-body observables.

%========================================================================================
% B-B-C system
%========================================================================================

\begin{figure}
\centering
\includegraphics[width=\linewidth]{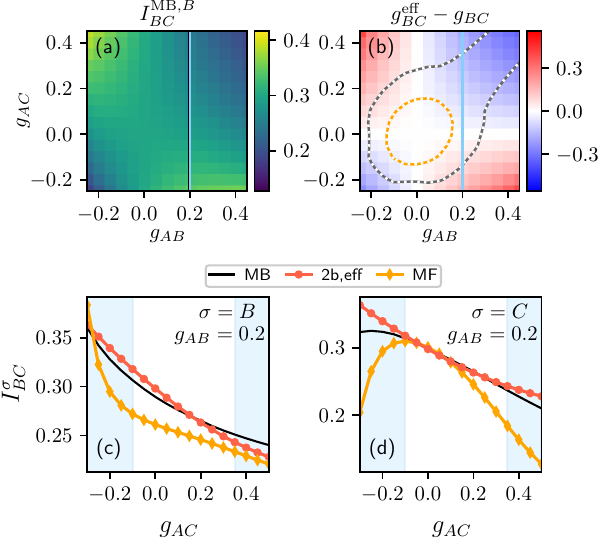}
\caption{(a) Integrated one-body density of the impurity $B$ (in the energetically higher well) as a function of the impurity-medium coupling strengths $g_{AB}$ and $g_{AC}$. (b) Effective two-body interaction strength (subtracting $g_{BC}$) obtained from an effective two-body model [cf. Eq. (\ref{eq:eff_ham_BB})]. Integrated density of the (c) $B$ and (d) $C$ impurities for varying $g_{AC}$ and fixed $g_{AB}=0.2$ [see the vertical blue solid lines in panels (a) and (b)] obtained within the many-body (MB), mean-field (MF) and effective two-body model (2b,eff). The encircled parametric regions in panel (b) where $\mathcal{E}_{BC}^{\mathrm{MF}}<0.03$ (orange-white dotted line) and where $\mathcal{E}_{BC}^{\mathrm{2b,eff}}<0.03$ (gray-white dotted line) signify an extended range of validity of the effective two-body model compared to the mean-field results. The blue shaded areas in panels (c), (d) denote $\mathcal{E}_{BC}^{\mathrm{2b,eff}}>0.03$.
In all panels, the system consists of a weakly-interacting bosonic ultracold gas in a ring potential coupled to two impurities $B$ and $C$ interacting with $g_{BC}=0.1$.}
\label{fig:BC_results}
\end{figure}

%-------------------------------------------------------------------------------------
\section{Induced interactions in three-component mixtures with impurities}
\label{three_component_induced}
%-------------------------------------------------------------------------------------

In the following, we examine the impact of mediated interactions on the behavior of two distinguishable impurities and two bosonic impurities plus one distinguishable impurity. Such a scenario presumes a three-component ultra-cold mixture. 
In our case, this consists of a medium $A$ confined to a ring and two distinct impurity species $B$ and $C$ trapped in a double-well potential. The many-body Hamiltonian of Eq. (\ref{eq:MB_hamiltionian_AB}) is readily extended to the three-component case:
\begin{align}
\hat{H} = \hat{H}_A + \hat{H}_B + \hat{H}_C + \hat{H}_{AB} + \hat{H}_{AC}.
\label{eq:MB_hamiltionian_ABC}
\end{align}
For simplicity, we consider a mass-balanced system, $m_A=m_B=m_C=1$, whilst  the impurities interact with $g_{BB}=g_{BC}=0.1$.

Analogously to the setup containing  indistinguishable bosonic impurities discussed in Sections \ref{sec:two_impurities_BB} and~\ref{sec:three_impurities_BBB}, our aim is to examine the impact of the mediated interactions between the impurities in a double-well potential. This will be again achieved by analyzing the effect of interactions on the one-body densities within the many-body method and the suitable effective model.
A key difference from the previously studied scenarios is the possibility to independently tune the impurity-medium interaction strengths. This gives rise to a substantial change in the character of the induced interactions. In particular, coupling one impurity attractively to the bath, while the other repels it, induces  repulsive interactions between the impurities as was also argued in Refs.~\cite{schecter2014, reichert2019, brauneis2021, theel2024}.

\subsection{Two distinguishable impurities}
\label{sec:two_impurities_BC}

In the following, we consider two distinguishable impurities, $N_B=1$ and $N_C=1$.
We start by examining the integrated one-body density of the $B$-impurity, $I_{BC}^{\mathrm{MB},B}$, determined within the many-body approach, see Fig. \ref{fig:BC_results}(a). Comparing the region where $g_{AB}=g_{AC}=0$ with the upper left (lower right) corner of Fig. \ref{fig:BC_results}(a), we find an increase of the $B$ impurity population in the energetically higher double-well site, while the regions corresponding to the upper right (lower left) corner show a reduction of $I_{BC}^{\mathrm{MB},B}$.
This signals the presence of a mediated repulsive (attractive) interaction between the impurities characterized by $g_{AB}g_{AC}<0$ ($g_{AB}g_{AC}>0$) in agreement with Refs.~\cite{schecter2014, reichert2019, brauneis2021, theel2024}, see also the discussion below.

Following Section~\ref{sec:BB_eff_2b_model}, we quantify this induced interaction using an effective two-body model
\begin{align}
\hat{H}_{BC}^{\mathrm{eff}} = \hat{H}_{B}^{\mathrm{eff}} (x^B) + \hat{H}_{C}^{\mathrm{eff}} (x^C) + g_{BC}^{\mathrm{eff}} \delta(x^B - x^C),
\label{eq:eff_ham_BC}
\end{align}
where the effective one-body Hamiltonian, $\hat{H}_{\sigma}^{\mathrm{eff}}$ (with  $\sigma=B,C$), is constructed similarly to the one described in Eq. (\ref{eq:eff_ham_B}). 
The effective two-body interaction strength $g_{BC}^{\mathrm{eff}}$ is tuned such that the ground-state energy of this effective model, $E_{BC}^{\mathrm{eff}}$, coincides with the expansion that contains one-body energies, i.e., $E_{BC}^{\mathrm{eff}} \overset{!}{=} E_B^\mathrm{pol} + E_C^\mathrm{pol} + E_{BC}^\mathrm{pol}$. Here, $ E_{BC}^\mathrm{pol}$ parametrizes correlations between two impurities.
Figure~\ref{fig:BC_results}(b) presents the effective two-body interaction strength $g_{BC}^{\mathrm{eff}}$ along the parametric $g_{AB}-g_{BC}$ plane. 
As it can be seen, the prediction made from the behavior of $I_{BC}^{\mathrm{MB},B}$ shown in Fig. \ref{fig:BC_results}(a), i.e., that $g_{BC}^{\mathrm{eff}} - g_{BC} \sim - g_{AB} g_{AC}$, is readily confirmed.

To judge the quality of the applied effective model, in Figs. \ref{fig:BC_results}(c) and (d) we analyze the integrated densities of impurities $B$ and $C$, respectively, for fixed $g_{AB}=0.2$ and variable $g_{AC}$. We find reasonable agreement in the weakly- to intermediate-interacting regions marked by the blue shaded areas in panels (c) and (d). The many-body and effective model predictions deviate from the corresponding mean-field calculations, in some cases even qualitatively. This observation is consistent with the case of two $B$ impurities, see Fig.~\ref{fig:BB_results_fit}.

To quantify the deviations of the effective two-body model from the many-body results, we sum over the relative differences of the integrated densities 
\begin{equation}
\mathcal{E}_{BC}^{\mathrm{2b,eff}} = \frac{\delta I_{BC}^{\mathrm{2b,eff},B} + \delta I_{BC}^{\mathrm{2b,eff},C}}{2},
\end{equation}
with $\delta I_{BC}^{\mathrm{2b,eff},\sigma} = |I_{BC}^{\mathrm{MB}, \sigma} - I_{BC}^{\mathrm{2b,eff}, \sigma}|/I_{BC}^{\mathrm{MB}, \sigma}$.  We estimate the deviations between the many-body and mean-field results in an analogous way, i.e., with $\mathcal{E}_{BC}^{\mathrm{MF}}$ involving  $\delta I_{BC}^{\mathrm{MF}, \sigma}$.
In Fig. \ref{fig:BC_results} (b) depicting $g_{BC}^{\mathrm{eff}}$, we encircle the parametric regions where $\mathcal{E}_{BC}^{\mathrm{2b,eff}}<0.03$ with a gray-white dotted line, while the ones characterized by $\mathcal{E}_{BC}^{\mathrm{MF}}<0.03$ lie within the orange-white dotted line.
It becomes apparent that the region of validity of the effective model is larger than that of the mean-field approximation. For the three-impurity system studied in the next section, we consider the parametric region where $\mathcal{E}_{BC}^{\mathrm{2b,eff}}<0.03$ as the area where the two-body model provides a reliable starting point for investigation.

\subsection{Two bosonic impurities and one distinguishable one}
\label{sec:two_impurities_BBC}

Finally, we consider a three-component mixture consisting of a bath $A$ coupled to two bosonic $B$ and one $C$ impurities. Similarly to the case of three indistinguishable  impurities (Section \ref{sec:three_impurities_BBB}), our goal here is to identify qualitative features of effective three-body interactions among the impurities, mediated by the bath. However, the important advantage of the present setting is that it encompasses two adjustable interaction parameters and not just one. This means that, besides $g_{AB}$, it is possible to also tune the interaction strength $g_{AC}$. Below, we analyze weakly-interacting impurities with $g_{BB}=g_{BC}=0.1$.

\begin{figure}
\centering
\includegraphics[width=\linewidth]{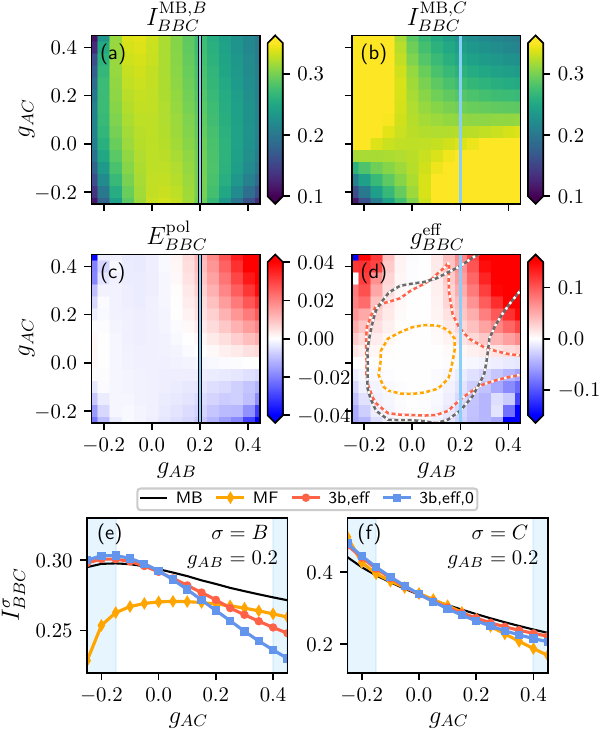}
\caption{Integrated one-body density (across the energetically higher well) of the impurities (a) $B$ and (b) $C$ as a function of the impurity-medium coupling strengths $g_{AB}$ and $g_{AC}$. (c) Three-polaron energy calculated from Eq. (\ref{eq:Epol_BBC}). (d) Effective three-body interaction strength in the parametric plane $g_{AB}-g_{AC}$, computed within the effective three-body model described by Eq. (\ref{eq:eff_ham_BBC}). Integrated density of impurities (e) $B$ and (f) $C$ for varying $g_{AC}$ and fixed $g_{AB}=0.2$ (see also  blue lines in panels (a)-(d)) obtained within the many-body (MB), the mean-field (MF) and the effective three-body model with (3b,eff) and without (3b,eff,0) the three-body interaction term. In panel (d), we encircle the regions where $\mathcal{E}_{BBC}^{\mathrm{MF}}<0.03$ (orange-white dotted line) and $\mathcal{E}_{BBC}^{\mathrm{3b,eff}}<0.03$ (red-white dotted line). The region encircled by the gray dashed line denotes the range of applicability of the effective two-body model, where $\mathcal{E}_{BBC}^{\mathrm{2b,eff}}<0.03$ holds. The blue shaded areas in (e) and (f) mark $\mathcal{E}_{BBC}^{\mathrm{2b,eff}}> 0.03$.
The three-component system comprises of a weakly-interacting bosonic ultracold gas on a ring potential coupled to two $B$-impurities and one $C$-impurity with $g_{BB}=g_{BC}=0.1$.}
\label{fig:BBC_results}
\end{figure}

As a first step, we examine the integrated one-body densities of the $B$ and $C$ impurities, see Figs.~\ref{fig:BBC_results}(a) and (b). A prominent difference between $I^{\mathrm{MB},B}_{BBC}$ and $I^{\mathrm{MB},C}_{BBC}$ is that the latter is larger in amplitude for $g_{AB} g_{AC} <0$. This behavior does not necessarily allude to a three-body effect, since in these parameter regions we expect an induced two-body repulsion~\cite{reichert2019,brauneis2021, theel2024} among the $B$ and $C$ impurities. Repulsion implies that the two $B$ impurities push the single $C$-impurity into the energetically higher site and hence $I^{\mathrm{MB},C}_{BBC}$ becomes larger. 
Therefore, in order to grasp the effects of a mediated three-body interaction on the one-body density, we need an effective three-body model.

We construct such a model by decomposing the energy of the system with three impurities $E_{BBC}^{(3)}$ in analogy to Eq.~(\ref{eq:Epol_BBB}):
\begin{align}
E_{BBC}^{(3)}  = E_A^{(0)} + 2E_B^{\mathrm{pol}} + E_C^{\mathrm{pol}} + E_{BB}^{\mathrm{pol}} + 2E_{BC}^{\mathrm{pol}} + E_{BBC}^{\mathrm{pol}}.
\label{eq:Epol_BBC}
\end{align}
In this expression, $E_B^{\mathrm{pol}}$, $E_C^{\mathrm{pol}}$ are the energies of a single dressed impurity; $E_{BB}^{\mathrm{pol}}$ and $E_{BC}^{\mathrm{pol}}$ describe the impurity-impurity and $E_{BBC}^{\mathrm{pol}}$ the three-impurity energies.
These energies are computed using the recipe of Section \ref{sec:three_impurities_BBB}. In particular, the single impurity energies are fitted to the effective one-body Hamiltonians $\hat{H}_B^{\mathrm{eff}}$ and $\hat{H}_C^{\mathrm{eff}}$. The energies $E_{BB}^{\mathrm{pol}}$ and $E_{BC}^{\mathrm{pol}}$ as well as the parameters $g_{BB}^{\mathrm{eff}}$, $g_{BC}^{\mathrm{eff}}$ are obtained from the corresponding  two-body models. 

The three-impurity effective Hamiltonian reads
\begin{align}
\hat{H}_{BBC}^{\mathrm{eff}} =& \sum_{i=1}^{2} \hat{H}_B^{\mathrm{eff}}(x_i^B) + \hat{H}_C^{\mathrm{eff}}(x^C) \nonumber \\
&+ g_{BB}^{\mathrm{eff}} \delta(x_1^B-x_2^B) + g_{BC}^{\mathrm{eff}} \sum_{i=1}^{2} \delta(x_i^B-x^C) \nonumber \\
&+ g_{BBC}^{\mathrm{eff}} \delta(x_1^B-x_2^B)\delta(x_2^B-x^C)
\label{eq:eff_ham_BBC}.
\end{align}
The last term describes the induced three-impurity correlations. The respective three-body interaction strength is tuned so that the ground-state energy of the effective model, $E_{BBC}^{\mathrm{eff}}$, matches the right-hand side of Eq. (\ref{eq:Epol_BBC}).

In Figures \ref{fig:BBC_results}(c) and (d) we present the three-polaron energy and the effective three-body interaction strength respectively, as a function of $g_{AB}$ and $g_{AC}$. It can be discerned that the sign of the three-body interaction strength qualitatively obeys the relation $g_{BBC}^{\mathrm{eff}} \sim g_{AB}^2 g_{AC}$. This property can be understood as a generalization of the case with three indistinguishable bosonic impurities where the sign of the three-body interaction strength follows $g_{BBB}^{\mathrm{eff}} \sim g_{AB}^3$. Note that the sign of the energy in Fig.~\ref{fig:BBC_results}(c) does not follow the simple prescription $g_{AB}^2 g_{AC}$. Although the exact origin of this behavior is not clear, our interpretation is the following. In general, an accurate description of the system with three impurities (counterintuitively) requires a modification of the two-body interaction, see the last term in Eq.~(\ref{eq:delta3}). This interpretation is in agreement with our results in Fig. \ref{fig:BBB_results_fit} where the three-impurity model does not describe the data accurately. 

To explicate the impact of $g_{BBC}^{\mathrm{eff}}$ on the accuracy of the effective three-body model, in Figs. \ref{fig:BBC_results}~(e) and~(f), we compare its predictions in the presence ($I_{BBC}^{\mathrm{3b,eff}, \sigma}$) and absence ($I_{BBC}^{\mathrm{3b,eff,0}, \sigma}$) of the three-body term for the integrated one-body density to the many-body prediction ($I_{BBC}^{\mathrm{MB}, \sigma}$), where $\sigma=B,C$. The results are shown for varying $g_{AB}$ and fixed $g_{AB}=0.2$. In view of the integrated densities of species $B$ and $C$, it can be seen that including the three-body term always improves the model's prediction for the integrated one-body density, provided that the effective two-body model is accurate. The parameter region where the two-body model is judged to be adequate corresponds to the region defined by $\mathcal{E}_{BBC}^{\mathrm{2b,eff}}= (\delta I_{BC}^{\mathrm{2b,eff},B} + \delta I_{BC}^{\mathrm{2b,eff},C} + \delta I_{BB}^{\mathrm{2b,eff},B})/3<0.03$, see gray-white dotted line in Fig. \ref{fig:BBC_results}(d). The regions where the two-body model looses its accuracy are marked by blue shaded areas in Figs. \ref{fig:BBC_results}(e) and (f), where $\mathcal{E}_{BBC}^{\mathrm{2b,eff}}>0.03$ holds.
The region where the effective three-body model produces qualitatively good predictions for the integrated one-body densities corresponds to the region where $\mathcal{E}_{BBC}^{\mathrm{3b,eff}} = (\delta I_{BBC}^{\mathrm{3b,eff},B} + \delta I_{BBC}^{\mathrm{3b,eff},C})/2 <0.03$, being encircled with a red-white dotted line in Fig. \ref{fig:BBC_results}(d).

Finally, in order to further reveal the role of correlations we compare the many-body with the mean-field results and estimate the region where $\mathcal{E}_{BBC}^{\mathrm{MF}} = (\delta I_{BBC}^{\mathrm{MF},B} + \delta I_{BBC}^{\mathrm{MF},C})/2 <0.03$, see the  parametric region surrounded by the orange-white dotted line in Fig.~\ref{fig:BBC_results}(d). Similar to the two-impurity case (Section \ref{sec:two_impurities_BC}), the region where the mean-field predictions are in good agreement with the many-body results for three impurities is limited to weak impurity-medium interactions $g_{AB}$ and $g_{AC}$.

\section{Conclusions and perspectives}
\label{conclusions}

We have studied the emergence of two- and three-body mediated interactions for a few impurity atoms. These atoms are trapped by a tilted double-well potential and immersed in a bosonic host, which is confined to a one-dimensional ring trap. 
This simple setup is suggested to be a prototype for detecting effects of induced interactions.  
To achieve a comprehensive description of the underlying induced interactions, two- and three-component mixture settings have been considered. Particular attention has been given to how the impurity-medium coupling strength influences the imbalance in impurity population at different sites of the double-well potential. 
To elucidate the sign and strength of induced interactions, effective two- and three-body models have been devised according to which the mediated interaction between the impurities is approximated by effective two- and three-body contact potentials. 

The associated effective interaction strengths, determined by fitting to the respective polaron energies of the many-body system, are found to be either attractive or repulsive depending on the impurity-medium coupling. 
It is showcased that the two-body model predictions are in a reasonable agreement with the results obtained from an \textit{ab initio} many-body approach. At the same time, the three-body effective models replicate many-body calculations only qualitatively, which we interpret as a general feature of the simplest three-polaron models. 
Additionally, we have compared the many-body results with relevant mean-field calculations, highlighting deviations in energies and densities.
These discrepancies naturally originate from the absence of interspecies correlations that are neglected within the mean-field framework. 
For instance, it is known that the mean-field approach results in a faster localization of the impurity~\cite{timmermans1998,Zschetzsche2024,Breu2025}, in our case, at the energetically lower double-well site.

There are a number of possible follow-up studies that we find worth-pursuing. 
First, it is important to investigate the effect of temperature on our results.
In general, one can expect that tight trapping of the impurities can help to circumvent one of the main problems in observing induced interactions, namely, the temperature of the bath. Indeed, trapping minimizes the excitation energy of the impurity, enhancing the effect of weak interactions. Second, it appears interesting to study
mixtures with larger atom numbers to testify the robustness of the effective interactions. As we have shown the simplest effective models fail to provide quantitatively accurate results already for three impurities.
In a similar vein, another perspective is to construct methods that are able to operate within the interaction regime of stronger intercomponent attractions where the strength of the effective interactions may be enhanced.

The generalization of the effective models to higher spatial dimensions as well as to Fermi/Bose systems containing fermionic or bosonic impurities can be a non-trivial extension~\cite{tajima2021,Baroni2023}.
 Similarly, it might be interesting to understand the role of three-body induced interactions for charged impurities~\cite{Astrakharchik2023}.
Finally, we note that the double-well potential without a tilt may contain information about induced impurity-impurity interactions in a two-body correlation function, see, e.g., Ref.~\cite{theel2022}. It appears interesting to study the impact of the induced three-body interaction in such a setting assuming a fine-tuned regime where the effect of the two-body induced potential is balanced by two-body free-space interactions.
It seems natural to design a suitable radiofrequency spectroscopy scheme, which 
would allow to reveal properties of the dressed states such as lifetime, residue, effective mass and importantly identify effects of induced two- and three-body interactions.

\section*{Acknowledgements}

S.I.M. acknowledges support from the University of Missouri Science and Technology, Department of Physics, in the framework of a startup fund. A.G.V. has been supported in part by the Danish National Research Foundation through the Center of Excellence ``CCQ'' (DNRF152). 
D.D. and P.S. acknowledge funding by the Cluster of Excellence “Advanced Imaging of Matter” of the Deutsche Forschungsgemeinschaft (DFG) - EXC 2056 - project ID 390715994.

\appendix

%-------------------------------------------------------------------------------
\section{Exact diagonalization method}
\label{app:effective_model}
%-------------------------------------------------------------------------------

In the following, we describe our approach to numerically solve the effective two and three-body models, given by Eqs. (\ref{eq:eff_ham_BB}), (\ref{eq:eff_ham_BBB}), (\ref{eq:eff_ham_BC}) and (\ref{eq:eff_ham_BBC}). We employ an exact diagonalization method in which the two- and three-body Hamiltonian matrix is constructed using corresponding two- and three-body basis states.
These basis states are used to build a wave function (see below) which is inserted in the time-independent Schr\"{o}dinger equation leading to
a set of coupled linear equations,
\begin{align}
\hat{H}^{(N_B),\mathrm{eff}} \mathbf{C}_n  = E_n \mathbf{C}_n. 
\label{eq:eff_SGL}
\end{align}
Here, $\mathbf{C}_n$ represents the coefficient vector to the $n$-th eigenenergy. 
After diagonalization it is possible to assess the ground state energy and wave function.
Exact diagonalization is a versatile method which has been employed, for instance, to systems consisting of few-body Bose mixtures \cite{AnhTai2024, anh-tai2025} or impurities coupled to a bosonic bath in a lattice \cite{Isaule2024}.
Below, we outline the construction of the aforementioned basis-states for either three indistinguishable bosonic impurities or two bosonic and one distinguishable impurities.

\subsection{Many-body basis for indistinguishable bosonic impurities}

Let us assume $N_B=2,3$ bosonic impurities. The statistical properties of the impurities are treated by expanding their wave function, $\Psi_{BB(B)}^{\mathrm{eff}}$, in terms of bosonic number states,
\begin{align}
|\Psi_{BB(B)}^{\mathrm{eff}} \rangle = \sum_{i=1}^{\mathcal{D}_B} C_i |\Vec{n}_i^B\rangle,
\label{eq:eff_wfct_B}
\end{align}
where $\Vec{n}^B_i=(n_1, \dots,n_{d_B})$. The latter denotes the occupation distribution of $N_B$ impurities over $d_B$ SPFs, while simultaneously satisfying the constraint $\sum_i n_i = N_B$. This leads to a total number of $\mathcal{D}_B=(d_B+N_B-1)!/[N_B!(d_B-1)!]$ number states. 
We ensure convergence of the applied method by providing a sufficient number of SPFs from which the number-state basis is formed.
As SPFs we choose the first $d_B=8$ energetically lowest eigenstates $\{\varphi_i^B(x)\}_{i=1}^{d_B}$ of the single-particle Hamiltonian $\hat{h}_B^{(1)}$ describing one atom in a 1D tilted double-well potential [Eq.~(\ref{bath_Hamilt})].

\subsection{Distinguishable bosonic impurities}

Next, we turn to a setting containing $N_B=1,2$ bosonic impurities and another distinguishable impurity of species $C$, i.e., $N_C=1$. The corresponding wave function ansatz, $\Psi_{B(B)C}^{\mathrm{eff}}$, has the form,
\begin{align}
|\Psi_{B(B)C}^{\mathrm{eff}} \rangle = \sum_{i=1}^{\mathcal{D}_B}\sum_{j=1}^{d_C} C_{ij} |\Vec{n}_i^B\rangle\otimes |\varphi_j^C\rangle,  
\end{align}
where the impurity $C$ is described in terms of the eigenfunctions $\{\varphi_i^C(x)\}_{i=1}^{d_C}$ of the single-particle Hamiltonian $\hat{h}_C^{(1)}$. Analogously to Eq. (\ref{eq:eff_wfct_B}), the $B$-impurities are expanded in terms of number states, which reduce, in the case of $N_B=1$, to a one-body basis. Finally, to solve this two- or three-body system, we evaluate the respective linear equation system given by the Schr\"{o}dinger equation [cf. Eq. (\ref{eq:eff_SGL})]. Throughout this work we consider $d_B=d_C=8$ SPFs, which ensure the convergence of our simulations.

%-------------------------------------------------------------------------------
\section{A single impurity coupled to the bosonic host}
\label{app:single_impurity_effmass}
%-------------------------------------------------------------------------------

\begin{figure}
\centering
\includegraphics[width=\linewidth]{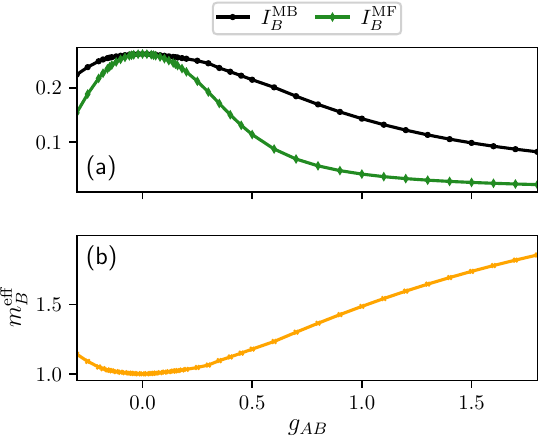}
\caption{(a) Integrated one-body densities $I_B^{\mathrm{MB}}$ and $I_B^{\mathrm{MF}}$ computed within the many-body and mean-field approaches, respectively [see also Eq. (\ref{eq:int_obd})]. Here, a single impurity is  trapped in a tilted double-well potential. The impurity is coupled via a contact interaction potential of strength $g_{AB}$ to a weakly interacting bath ($g_{AA}=0.1$) residing on a ring potential and containing $N_A=12$ atoms. (b) Effective mass of the impurity extracted by fitting the integrated one-body density of an effective one-body model [Eq. (\ref{eq:eff_ham_B_effmass})] to $I_B^{\mathrm{MB}}$.}
\label{fig:B_results}
\end{figure}

For completeness, we investigate the scenario of a majority species $A$ confined in a ring potential and being coupled to a single impurity $B$ trapped in a tilted double-well potential. 
This system allows for explicit comparisons with the two (or three) impurity settings ultimately hinting towards the necessity to account for induced interactions. 

The respective integrated one-body densities $I_B^{\mathrm{MB}}$ (over the energetically higher site) obtained from the many-body approach as a function of $g_{AB}$ are presented in Fig. \ref{fig:B_results}(a). As it can be seen, the density portion in the energetically elevated double-well site decreases as $|g_{AB}|$ increases indicating a localization of the impurity at the energetically lower site. 
Recall that a similar trend of the integrated density takes place for two or three bosonic impurities, see e.g. Fig.~\ref{fig:BB_results_fit}(b) and Fig.~\ref{fig:BBB_results_fit}(b).
Comparing the behaviors of the one- and two-impurity cases, $I_B^{\mathrm{MB}}$ and $I_{BB}^{\mathrm{MB}}$, respectively, we find that the localization trend is stronger in the case of two non-interacting impurities (not shown) than the one of a single impurity. 
This suggests the presence of an induced attractive interaction.
Moreover, this behavior is compared to $I_B^{\mathrm{MF}}$, i.e., the integrated density obtained within a mean-field approximation. 
It is found that in general $I_B^{\mathrm{MF}}<I_B^{\mathrm{MB}}$ implying that correlations impede (but not eventually prevent) the impurity's localization Ref.~\cite{Zschetzsche2024,Breu2025}).

Another important aspect of the dressed impurity that has not been estimated thus far is its effective mass. This may support the localization tendency of the impurities for varying interactions but also improve the agreement of the effective models with the many-body computations (see also Appendix~\ref{app:two_impurities_effmass}).
For this reason, we construct an effective one-body description which in fact has been intensively used before and argued to provide an adequate approximation both for the static but also the dynamical properties of a single impurity~\cite{mistakidis2019a,mistakidis2020b}. It reads 
\begin{align}
\hat{H}_B^{\mathrm{eff'}} = \epsilon_B^{\mathrm{pol}'} -\frac{\hbar^2}{2 m_B^{\mathrm{eff}}} \partial_x^2 + V_B(x),
\label{eq:eff_ham_B_effmass}
\end{align}
where $V_B(x)$ denotes the tilted double-well potential [Eq. (\ref{eq:double_well_potential})]. The energy difference $\epsilon_B^{\mathrm{pol}'}$ is chosen such that the ground-state energy of the effective model $\hat{H}_B^{\mathrm{eff'}}$ matches the polaron energy $E_B^{\mathrm{pol}}$ [Eq. (\ref{eq:Epol_B})].

The crucial difference with the model described by Eq. (\ref{eq:eff_ham_B}) is the presence of the effective mass, $m_B^{\mathrm{eff}}$. The latter is determined by varying $m_B^{\mathrm{eff}}$ such that the integrated one-body density of the effective model ($I_B^{\mathrm{1b,eff}}$) coincides with the many-body result ($I_B^{\mathrm{MB}}$). 
This is achieved by minimizing the cost function $|I_B^{\mathrm{MB}} - I_B^{\mathrm{1b,eff}}|^2$.
An overall accuracy of $\sim10^{-10}$ is ensured within our simulations. In Fig. \ref{fig:B_results}(b) we present the results for the effective mass as a function of $g_{AB}$. As shown, increasing $|g_{AB}|$ leads to a larger effective mass which is in accordance with the observed localization behavior of the impurities in the main text. Recall that in the main text we argued that considering the bare impurity mass in the effective models improves the agreement with the many-body results. Below, in Appendix \ref{app:two_impurities_effmass}, we exemplify the impact of the effective mass on the outcome of the effective two-body model.

%-------------------------------------------------------------------------------
\section{Impact of the effective mass on the two-impurity behavior}
\label{app:two_impurities_effmass}
%-------------------------------------------------------------------------------

\begin{figure}
\centering
\includegraphics[width=\linewidth]{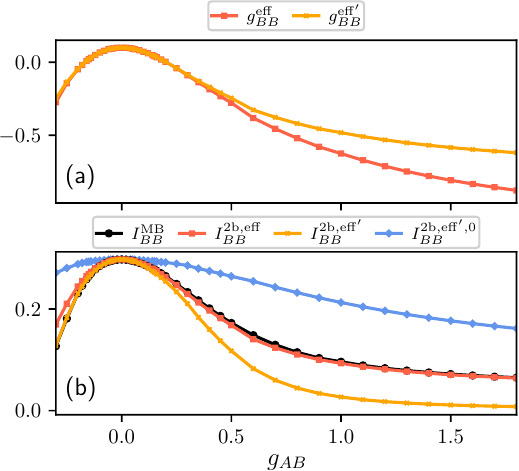}
\caption{(a) Effective two-body interaction strengths $g_{BB}^{\mathrm{eff}}$ and $g_{BB}^{\mathrm{eff}'}$ extracted from the two-body models given by Eq. (\ref{eq:eff_ham_BB}) and Eq. (\ref{eq:eff_ham_BB_effmass}), respectively. (b) Integrated one-body density obtained within the many-body approach ($I_{BB}^{\mathrm{MB}}$) and with an effective two-body model which considers the bare impurity mass [$I_{BB}^{\mathrm{2b,eff}}$, Eq. (\ref{eq:eff_ham_BB})] and an effective mass [$I_{BB}^{\mathrm{2b,eff'}}$, Eq.~(\ref{eq:eff_ham_BB_effmass})]. The effect of the induced interaction is highlighted by a comparison to $I_{BB}^{\mathrm{eff',0}}$ corresponding to an effective two-body model which includes the effective mass but neglects the induced interaction, i.e., in Eq.~(\ref{eq:eff_ham_BB_effmass}) we set $g_{BB}^{\mathrm{eff'}}\equiv g_{BB}=0.1$. We consider two interacting ($g_{BB}=0.1$) bosonic impurities, which are coupled to a bath consisting of $N_A=12$ bosons featuring $g_{AA}=0.1$.}
\label{fig:BB_results_effmass}
\end{figure}

Having at hand $m_B^{\mathrm{eff}}$ (see Appendix~\ref{app:single_impurity_effmass}) it is instructive to study its impact on the accuracy of the applied effective two-body model described in Section \ref{sec:BB_eff_2b_model}. Indeed, the effective mass may be utilized to construct a corresponding effective two-body model,
\begin{align}
\hat{H}_{BB}^{\rm{eff}'} =& \sum_{i=1}^{2} \hat{H}_{B}^{\rm{eff'}}(x_i) + g_{BB}^{\mathrm{eff'}}\delta(x_1-x_2),
\label{eq:eff_ham_BB_effmass}
\end{align}
where $\hat{H}_{B}^{\rm{eff'}}$ refers to the one-body Hamiltonian of Eq. (\ref{eq:eff_ham_B_effmass}). Similarly to the prescription followed in Section \ref{sec:BB_eff_2b_model}, the effective two-body interaction strength $g_{BB}^{\mathrm{eff'}}$ is estimated by demanding a matching of the ground-state energy of $\hat{H}_{BB}^{\rm{eff}'}$ with the energy $2E_B^{\mathrm{pol}} + E_{BB}^{\mathrm{pol}}$. In Fig. \ref{fig:BB_results_effmass}(a) we directly compare the effective two-body interaction strengths, $g_{BB}^{\mathrm{eff}}$ and $g_{BB}^{\mathrm{eff'}}$, corresponding to the two-body models without [Eq.~(\ref{eq:eff_ham_BB})] and with [Eq.~(\ref{eq:eff_ham_BB_effmass})] an effective mass, respectively. It is evident that for small impurity-medium interaction strengths, $g_{AB}$, of either sign the effective coupling parameters extracted for the aforementioned different models agree well with each other while deviating for larger repulsive $g_{AB}$. In particular, it appears that $g_{BB}^{\mathrm{eff'}}$ takes smaller absolute values than $g_{BB}^{\mathrm{eff}}$.

To judge the quality of the applied methods, we additionally compare the respective integrated one-body densities, see Fig. \ref{fig:BB_results_effmass}(b). For repulsive $g_{AB}$, we find that the best agreement compared to the many-body results, $I_{BB}^{\mathrm{MB}}$, occur for the integrated one-body density ($I_{BB}^{\mathrm{2b,eff}}$) calculated with the model described in Section \ref{sec:BB_eff_2b_model} where the effective mass has been neglected. On the other hand, the integrated density corresponding to the model in Eq. (\ref{eq:eff_ham_BB_effmass}), $I_{BB}^{\mathrm{2b,eff'}}$, underestimates the many-body results for repulsive $g_{AB}$. However, for attractive $g_{AB}$, including the effective mass results in a better agreement between $I_{BB}^{\mathrm{eff'}}$ and $I_{BB}^{\mathrm{MB}}$, while $I_{BB}^{\mathrm{eff}}$ overestimates the target results. 
This result is not straightforward, since a-priori one would expect an improvement of the model predictions when including the effective mass. Such an effect may emanate from different sources such as the rather complex potential landscape employed, or the fact that we operate far from the thermodynamic limit. To resolve this issue a careful analysis is required going even beyond the currently employed methods to exemplify the origin of this discrepancy which is left for future endeavors.  

To reveal the effect of the induced interaction $g_{BB}^{\mathrm{eff'}}$ on the integrated density, we additionally calculate the observable $I_{BB}^{2b,\mathrm{eff',0}}$, obtained from an effective two-body model which includes the effective mass but does not consider the effects of the mediated interactions, i.e., in Eq.~(\ref{eq:eff_ham_BB_effmass}) we set $g_{BB}^{\mathrm{eff'}}\equiv g_{BB}=0.1$. It can be readily seen from Fig. \ref{fig:BB_results_effmass}(b), that  $I_{BB}^{\mathrm{eff',0}}$ clearly deviates from $I_{BB}^{\mathrm{MB}}$ for $g_{AB}\neq0$ and in fact yields (at least within the considered parameter range) always larger values than $I_{BB}^{\mathrm{MB}}$. 
This indicates that, indeed, an attractive induced interaction strength is required to correctly capture the many-body results.

%-------------------------------------------------------------------------------
\section{Renormalization of the effective two-body interaction strength in a three-impurity setup}
\label{app:renorm_3body_int}
%-------------------------------------------------------------------------------
\begin{figure}
\centering
\includegraphics[width=\linewidth]{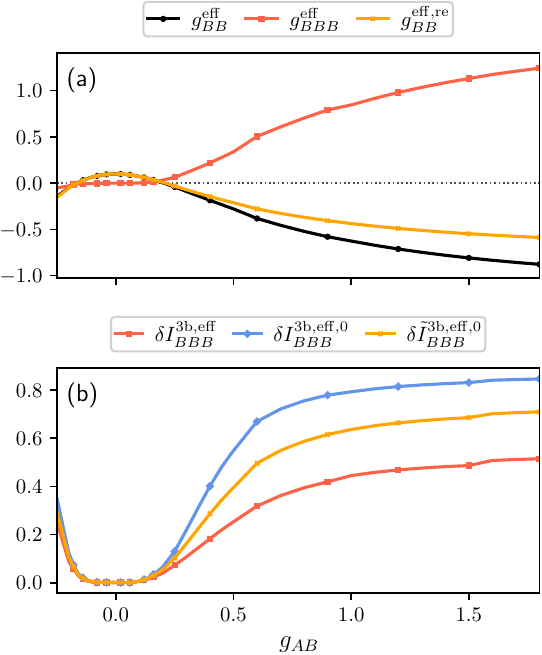}
\caption{
(a) Effective two-body (three-body) interaction strength $g_{BB}^{\mathrm{eff}}$ ($g_{BBB}^{\mathrm{eff}}$) obtained from the effective model in Eq.~(\ref{eq:eff_ham_BB}) [Eq.~(\ref{eq:eff_ham_BBB})] and the condition described by Eq.~(\ref{eq:energy_condition_BB}) [Eq.~(\ref{eq:energy_condition_BBB})]. These interaction strengths are compared to the renormalized, effective two-body interaction $g_{BB}^{\mathrm{eff,re}}$ obtained from $\tilde{\hat{H}}_{BBB}^{\mathrm{eff,0}}$ [cf. Eq.~(\ref{eq:eff_ham_BBB_v2})] and the condition given by Eq. (\ref{eq:energy_condition_BBB}).
(b) Relative differences $\delta I_{BBB}^{\mathrm{3b,eff}}$, $\delta I_{BBB}^{\mathrm{3b,eff,0}}$ and  $\delta \tilde{I}_{BBB}^{\mathrm{3b,eff,0}}$ between the integrated one-body density obtained with a specific effective three-body model and with a many-body method (see the text for details). Here, $N_A=12$, $N_B=2,3$ and $g_{AA}=g_{BB}=0.1$.
} 
\label{fig:BBB_fit_gBB_only}
\end{figure}

In this appendix, we pursue the question whether a three-body interaction term in the effective model in Eq. (\ref{eq:eff_ham_BBB}) is required for a better description of the one-body results or whether they can be captured by renormalizing the effective two-body interaction without a three-body term.

In order to answer this question, we neglect the three-body term in Eq. (\ref{eq:eff_ham_BBB}) by setting $g_{BBB}^{\mathrm{eff}}=0$ and replace $g_{BB}^{\mathrm{eff}}$ with a renormalized two-body interaction strength $g_{BB}^{\mathrm{eff,re}}$:
\begin{align}
    \tilde{\hat{H}}_{BBB}^{\mathrm{eff,0}} =& \sum_{i=1}^{3} H_{B}^{\mathrm{eff}}(x_i) +  g_{BB}^{\mathrm{eff,re}} \sum_{\substack{i,j=1 \\ i\neq j}}^{3} \delta(x_i - x_j).
    \label{eq:eff_ham_BBB_v2}
\end{align}
To determine  $g_{BB}^{\mathrm{eff,re}}$, we vary its value until the condition in Eq. (\ref{eq:energy_condition_BBB}) is fulfilled.

We apply this procedure to a system of $N_A=12$ weakly interacting majority particles coupled to three impurities interacting with $g_{AA}=g_{BB}=0.1$ as done in Sec. \ref{sec:three_impurities_BBB}.
In Fig.~\ref{fig:BBB_fit_gBB_only}(a) we present the modified two-body interaction strength $g_{BB}^{\mathrm{eff,re}}$ as a function of $g_{AB}$ together with the two- and three-body interaction strengths $g_{BB}^{\mathrm{eff}}$ and $g_{BBB}^{\mathrm{eff}}$ determined from the procedure discussed in Secs. \ref{sec:BB_eff_2b_model} and \ref{sec:three_impurities_BBB}. 

Inspecting Fig.~\ref{fig:BBB_fit_gBB_only}(a), we find at small impurity-medium interaction strengths $g_{AB}$, where the three-body interaction strength $g_{BBB}^{\mathrm{eff}}$ is close to zero, a good agreement between the modified two-body interaction strength $g_{BB}^{\mathrm{eff,re}}$ and $g_{BB}^{\mathrm{eff}}$. However, as we increase $g_{AB}$ to repulsive values, $g_{BB}^{\mathrm{eff,re}}$ becomes larger than $g_{BB}^{\mathrm{eff}}$. We interpret this by an increase of repulsive three-body interaction which the modified interaction strength $g_{BB}^{\mathrm{eff,re}}$ has to compensate for.

In order to judge the quality of the effective model of Eq.~(\ref{eq:eff_ham_BBB_v2}), we compare the respective integrated one-body density to the one obtained from the many-body method, $I_{BBB}^{\mathrm{MB}}$. For comparison, we also present results for the effective model from the main text. In particular, we calculate the relative difference, $\delta I_{BBB}^{\mathrm{3b,eff}} =|I_{BBB}^{\mathrm{MB}} - I_{BBB}^{\mathrm{3b,eff}} |/I_{BBB}^{\mathrm{MB}}$ and analogously $\delta I_{BBB}^{\mathrm{3b,eff,0}}$ and $\delta \tilde{I}_{BBB}^{\mathrm{3b,eff,0}}$. Here, $ I_{BBB}^{\mathrm{3b,eff}}$ and $I_{BBB}^{\mathrm{3b,eff,0}}$ represent the integrated one-body density obtained within the effective %three-body
model with and without the three-body term, respectively, see Fig. \ref{fig:BBB_results_fit}(b), and $\tilde{I}_{BBB}^{\mathrm{3b,eff,0}}$ denotes the integrated density of the effective model with renormalized two-body interaction strength %modified effective model %
$\tilde{\hat{H}}_{BBB}^{\mathrm{eff,0}}$ [Eq. (\ref {eq:eff_ham_BBB_v2})].
The relative differences are shown in Fig. \ref{fig:BBB_fit_gBB_only}(b) with respect to $g_{AB}$. As it can be readily seen, for large values of $g_{AB}$, we observe the best agreement for an effective 
%three-body
model with a three-body term ($\delta I_{BBB}^{\mathrm{3b,eff}}$), followed by the version %three-body model
with a modified two-body interaction strength ($\delta \tilde{I}_{BBB}^{\mathrm{3b,eff,0}}$). We find the largest deviation from the many-body results for the effective model which only includes the effective two-body interaction strength $\tilde{g}_{BB}^{\mathrm{eff}}$ ($\delta I_{BBB}^{\mathrm{3b,eff,0}}$).

%-------------------------------------------------------------------------------
\section{Correlation impact on one-body and two-body observables}
\label{app:impact_of_corr}
%-------------------------------------------------------------------------------

In this appendix, we employ different ansatzes for the many-body wave function in order to unravel the role of correlations on one- and two-body observables. The many-body wave function of a general three-component mixture (see also Section \ref{sec:ML-X}) is firstly expanded in terms of different $D_A$, $D_B$ and $D_C$ species functions. Setting $D_A = D_B = D_C = 1$, all intercomponent correlations are neglected rendering the total wave function [Eq.~(\ref{eq:psi_mlx})] a single product state where each species is represented by a single species wave function. Note that each species wave function is expanded in terms of different SPFs accounting for intracomponent correlations.
This approach is referred to as \textit{species mean-field} (sMF)~\cite{theel2024}. A step beyond this sMF ansatz is to allow entanglement formation solely between two of the species, while correlations with the third species are suppressed. This is accomplished, e.g. through $D_B=1$ and $D_A,D_C>1$, meaning that species $A$ and $C$ are correlated, experiencing a mean-field type potential from species $B$. We will dub this approach \textit{species mean-field of species $B$} (sMFB). 
Analogously, one can define sMFA and sMFC. 

As such it is possible to extract the impact of correlations between two different species on an arbitrary observable $\hat{O}$. For instance, to reveal the impact of correlations between species $A$ and $B$, we calculate,
\begin{align}
\Delta_{AB} = \langle \hat{O} \rangle_{\mathrm{sMFC}} - \langle \hat{O} \rangle_{\mathrm{sMF}},
\end{align}
where $\langle \hat{O} \rangle_{X}$ denotes the expectation value of $\hat{O}$ using the method $X=\textrm{MB}, \textrm{sMF}, \textrm{sMFA}, \dots$. In a next step, we can decompose the expectation value calculated within the full many-body method, i.e. $\langle \hat{O}\rangle_{\mathrm{MB}}$, in terms of contributions of different correlation orders as follows,
\begin{align}
\langle \hat{O}\rangle_{\mathrm{MB}} = \langle \hat{O}\rangle_{\mathrm{sMF}} + \Delta_{\mathrm{2 spec}} + \Delta_{\mathrm{3 spec}}.
\label{eq:obs_MB_2corr_3corr}
\end{align}
Here, the expectation value $\langle \hat{O}\rangle_{\mathrm{MB}}$, which includes all relevant interspecies correlations, splits into a species mean-field part, $\langle \hat{O}\rangle_{\mathrm{sMF}}$, as well as a second- and third-order correlation term, $\Delta_{\mathrm{2 spec}}=\Delta_{AB} + \Delta_{AC} + \Delta_{BC}$ and $\Delta_{\mathrm{3 spec}}$. Specifically, we calculate $\Delta_{\mathrm{3 spec}}$ by subtracting the species mean-field and the second-order contribution from the many-body result $\langle \hat{O}\rangle_{\mathrm{MB}}$. More detailed discussions about such a decomposition can be found in Ref.~\cite{theel2024}.

\begin{figure}
\centering
\includegraphics[width=\linewidth]{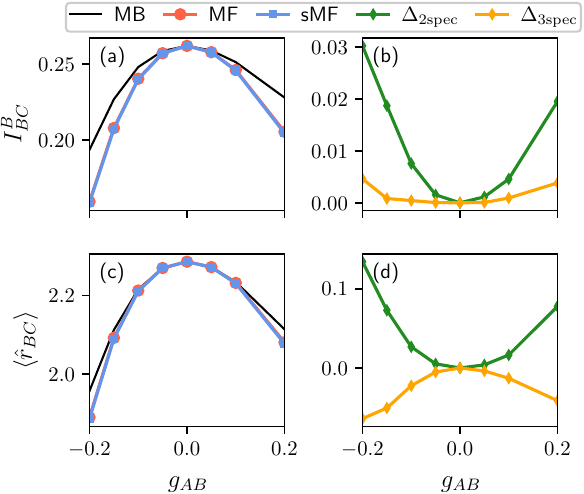}
\caption{(a) Integrated one-body density of impurity $B$, $I_{BC}^{B}$ [Eq. (\ref{eq:int_obd})], obtained within the full many-body (MB), mean-field (MF) and species mean-field (sMF) ansatz. (b) Contributions of the second- and third-order correlations on $I_{BC}^{B}$, see Eq. (\ref{eq:obs_MB_2corr_3corr}). Impurities' relative distance, $\langle \hat{r}_{BC}\rangle$, by (c) using different numerical approaches (see legend) and (d) distinguishing second- and third-order correlation effects. 
We consider two distinguishable impurities $B$ and $C$ trapped in a tilted double-well potential coupled to a weakly interacting majority species $A$ on a ring potential with $g_{BC}=0$ and $g_{AB}=g_{AC}$.}
\label{fig:impact_correlation}
\end{figure}

Below, we focus on the setup of Section~\ref{sec:two_impurities_BC}, i.e., two distinguishable impurities $B$ and $C$ coupled to a bosonic majority species $A$ where $g_{BC}=0$. 
The integrated one-body density of species $B$, $I_{BC}^{B}$ obtained within the many-body (MB), mean-field (MF) and sMF methods is presented in Fig.~\ref{fig:impact_correlation}(a) as a function of $g_{AB}=g_{AC}$. As it can be seen, the MF and sMF predictions show a very good agreement indicating that the presence of correlations among the particles of species $A$ do not play a decisive role for the behavior of the integrated density $I_{BC}^{B}$.
Interestingly, a close comparison among the results obtained with the sMF and the MB approaches, unveils that the absence of interspecies correlations leads to a faster decrease of the density at the energetically higher double-well site. 
This clearly showcases that interspecies correlations hinder the impurities localization. 
In Figure \ref{fig:impact_correlation}(b) we resolve the discrepancy between the MB and sMF calculations in terms of two- and three-body effects. It turns out, that both contributions lead to an increase of the integrated density with the two-body ones naturally exhibiting the larger participation.

Next, we aim to generalize our observations by comparing correlation effects imprinted on two-body observables such as the impurities' relative distance $\langle \hat{r}_{BC}\rangle$~\cite{mistakidis2019, mistakidis2020a}.
Also inspection of this observable explicates that the presence of all correlations leads to an increase of the impurities relative distance, see Fig.~\ref{fig:impact_correlation}(c). However, when resolving the impact of correlations in terms of second- and third order terms, we find that only two-component correlations are associated with an increase of $\langle \hat{r}_{BC}\rangle$, while the third order contribution is negative, see Fig.~\ref{fig:impact_correlation}(d). 
This third-order mechanism is attributed to an induced effect, where the majority species $A$ mediates correlations between the non-interacting impurities $B$ and $C$. In particular, the shrinking of $\langle \hat{r}_{BC}\rangle$ can be interpreted as a correlation-induced attraction, which is in accordance with the effects reported in Refs.~\cite{chen2018, theel2024}.

%-------------------------------------------------------------------------------
\section{Impact of $N_A$ on the polaron energies and the integrated impurity density}
\label{app:scaling_behavior}
%-------------------------------------------------------------------------------

\begin{figure}
\centering
\includegraphics[width=\linewidth]{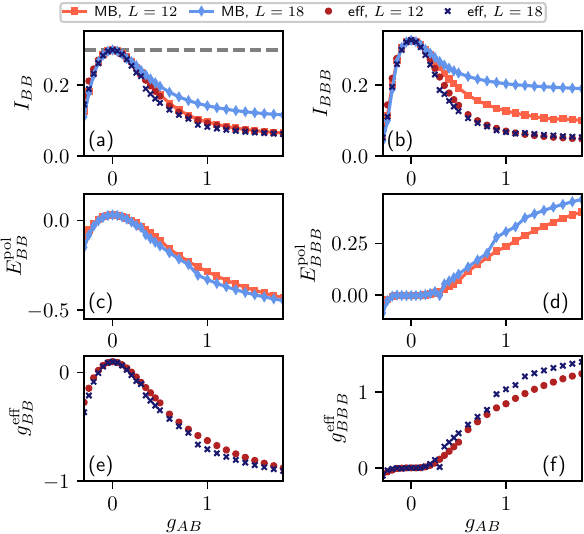}
\caption{
Integrated one-body densities $I_{BB}$ and $I_{BBB}$ [see Eq. (\ref{eq:int_obd})] for systems with (a) $N_B=2$ and (b) $N_B=3$ bosonic impurities interacting with $g_{BB}=0.1$. The results are obtained within a many-body approach (MB) and the respective effective models (eff), see Eqs. (\ref{eq:eff_ham_BB}) and (\ref{eq:eff_ham_BBB}).
The impurities are coupled to different majority species consisting of $N_A$ atoms confined in a ring potential of length $L$ and experiencing $g_{AA}=0.1$ (see legend).
The horizontal dashed line in panel (a) marks the population imbalance of the impurities, when they are decoupled from the majority species, i.e., when $g_{AB}=0$.
(c) Bipolaron [Eq.~(\ref{eq:Epol_BB})] and (d) three-polaron energies [Eq.~(\ref{eq:Epol_BBB})].
(e) Effective two- and (f) three-body interaction strength derived from the bipolaron and three-polaron energies.
}
\label{fig:NL_scaling}
\end{figure}

In this appendix, we elaborate on the impact of the majority species atom number $N_A$ (assuming a fixed density $N_A/L=1$) on the polaron energies and impurity densities.
In Figs.~\ref{fig:NL_scaling}(a) and (b), we show the integrated one-body density [Eq. (\ref{eq:int_obd})] for two and three impurities interacting with $g_{BB}=0.1$, respectively, for systems with ($N_A=12$, $L=12$) as well as ($N_A=18$, $L=18$) and obtained within the many-body approach (MB).
For $g_{AB}>0$, we always find that increasing the system size leads to a larger value of the integrated densities $I_{BB}$ and $I_{BBB}$. This behavior indicates that for larger system sizes ($N_A=18$ and $L=18$) the impurities feature less localization tendency at the energetically lower double-well site.

We attribute this behavior to a finite-size effect stemming from the majority species. In particular, we find for a ring size of $L=12$ and $N_A=12$ that the majority species exerts an increased pressure on the impurities facilitating their localization on the energetically lower double-well site. This pressure can be reduced by considering larger ring sizes, while keeping the ratio $ N_A /L$ constant. Increasing the system size of the majority species enhances the localization of the impurities leading to the observed behavior of $I_{BB}$ and $I_{BBB}$ [see solid lines in Figs.~\ref{fig:NL_scaling}(a) and (b)].

In Figs.~\ref{fig:NL_scaling}(c) and (d), we present the two-impurity as well as the three-impurity energies obtained with Eqs.~(\ref{eq:Epol_BB}) and (\ref{eq:Epol_BBB}), respectively.
Interestingly, we find that these observables are across the considered parameter range of $g_{AB}$ to a good degree independent of the system size, i.e., increasing the $L$ and $N_A$ while $N_A/L=\mathrm{const}$ does not significantly affect these energies.
As outlined in Sections \ref{sec:two_impurities_BB} and \ref{sec:three_impurities_BBB}, we can derive from these energies an effective two- and three-body interaction strength which is mediated between the impurities, i.e., $g_{BB}^{\mathrm{eff}}$ and $g_{BBB}^{\mathrm{eff}}$, shown in Figs.~\ref{fig:NL_scaling}(e) and (f). As the energies of the bipolaron and the three-polaron are largely independent of the system size, the same holds for the effective interaction strengths.
Likewise, inserting these effective interaction parameters into the two- and three-body models, described by  Eqs. (\ref{eq:eff_ham_BB}) and (\ref{eq:eff_ham_BBB}), it is possible to calculate $I_{BB}$ and $I_{BBB}$ [see circles and crosses in Figs.~\ref{fig:NL_scaling}(a) and (b)].
It can be seen that in both cases an overall decreasing behavior of the integrated densities occurs which, however, gradually deviates from the many-body predictions for increasing $g_{AB}$. Qualitative agreement is only observed for $I_{BB}$ in the ($L=12$, $N_A=12$) system.

To summarize, this appendix demonstrates the usefulness of the induced interactions for the description of the impurity system for sizes larger than the one studied in the main text.
Indeed, in all considered scenarios, a finite interspecies interaction strength $g_{AB} \neq 0$ leads to a decreased integrated density when comparing to the decoupled case ($g_{AB}=0$), see gray dashed line in Figure~\ref{fig:NL_scaling}(a). Induced interactions give us a simple tool 
for understanding and modeling this behavior.

%-------------------------------------------------------------------------------
\section{Diverging three-body interaction}
\label{app:diverging3-body}
%-------------------------------------------------------------------------------

Here, we will show that the three-body interaction used in the main text, see Eq.~\eqref{eq:eff_ham_BBB}, diverges logarithmically -- similar to the contact interaction in 2D (see, e.g., Refs.~\cite{rontani2017, nyeo2000regularization} and references therein). Therefore, we start with the following Hamiltonian of three particles in 1D with periodic boundary conditions:
\begin{equation}
H=-\frac{1}{2}\sum_{i=1}^{3}\frac{\partial^2}{\partial x_i^2}+g\delta(x_1-x_2)\delta(x_2-x_3).
\end{equation}
Since all interactions are translationally invariant, the total momentum of the system, $P$, is conserved. We can use this to eliminate one of the three coordinates by writing~\cite{Gross1962,Girardeau1983,mistakidis2019a}
\begin{equation}
    \Psi(x_2, x, y)=\Tilde{\Psi}(x, y)e^{iPx_2},
\end{equation}
where $x=x_1-x_2+\theta(x_2-x_1)$, $y=x_3-x_2+\theta(x_2-x_3)$ with $\theta(x)$ the Heaviside step function.
This approach leads to the following Hamiltonian 
\begin{equation}
    H=-\left(\frac{\partial^2}{\partial x^2}+\frac{\partial^2}{\partial y^2}+\frac{\partial}{\partial y}\frac{\partial}{\partial x}\right)-g\delta(x)\delta(y).
\end{equation}
Note that this Hamiltonian is reminiscent of the Hamiltonian of a particle with mass $m=2$ and a contact potential together with the additional mixed derivative term  $\frac{\partial}{\partial x}\frac{\partial}{\partial y}$.

Next, we solve the Schr\"odinger equation in momentum space to show that this Hamiltonian diverges logarithmically,
\begin{equation}
k^2\Phi(\Vec{k})+k_x k_y\Phi(\Vec{k})+g\Psi(0)=E\Phi(\Vec{k}).
\end{equation}
Rewriting this expression, inserting $\Psi(0)=\int\frac{d^2{k}'}{(2\pi)^2}\Phi(\Vec{k}')$ and integrating over both sides in $\vec{k}$ we can write:
\begin{equation}
\begin{split}
\int\frac{d^2{k}}{(2\pi)^2}\Phi(\Vec{k})&=\int\frac{d^2{k}}{(2\pi)^2}\frac{g}{-k^2-k_x k_y+E}\int\frac{d^2{k}'}{(2\pi)^2}\Phi(\Vec{k}')\\
1&=\int\frac{d^2{k}}{(2\pi)^2}\frac{g}{-k^2-k_x k_y+E}.
\end{split}
\end{equation}
With this expression, we can easily see that the integral is diverging logarithmically. We only run into problems for large values of $|k|$:
\begin{equation}
\int\frac{d^2{k}}{(2\pi)^2}\frac{g}{-k^2-k_x k_y+E}\approx\int\frac{d^2{k}}{(2\pi)^2}\frac{g}{-k^2-k_x k_y}. 
\end{equation}
Next, we introduce polar coordinates to write
\begin{equation}
\begin{split}
&\int\frac{d^2{k}}{(2\pi)^2}\frac{g}{-k^2-k_x k_y}=\int_0^\infty\frac{dk k}{(2\pi)^2} \\
& \times \int_0^{2\pi} d\phi\frac{g}{-k^2-k^2\sin{\phi}\cos{\phi}}=-\frac{g}{\sqrt{3}\pi}\int_0^\infty dk/k.
\end{split}
\end{equation}
It becomes apparent that this integral indeed diverges  logarithmically.

As mentioned briefly in the main text, this is however not problematic for our approach. Since we solve the Schr\"odinger equation numerically with a fixed cutoff, the latter regularizes the above integral. We renormalize it by matching the three-body interaction strength $g_{BBB}$ to the energy of the three quasi-particles. This approach is similar to common renormalization methods for Configuration-Interaction calculations in two-dimensional cold atom systems with contact interaction. In this context, the strength of the two-body interaction is determined by matching the bare value of the contact interaction strength to reproduce the same two-body ground state energy for each cutoff, see e.g. Ref.~\cite{rontani2017,Brauneis2025}.

\bibliography{literature_3imp.bib}

\end{document}